\documentclass[aps,pre,amsmath,amssymb,nobibnotes,superscriptaddress,
reprint]{revtex4-2}
\usepackage{graphicx}%
\usepackage{dcolumn}
\usepackage{bm}
\usepackage{chemformula}
\usepackage[utf8]{inputenc}
\usepackage{etoolbox}
\usepackage[autostyle]{csquotes}
\usepackage[tight]{subfigure}
\usepackage[unicode]{hyperref}
\hypersetup{colorlinks= true,
	citecolor= blue,
	linkcolor= red,
	urlcolor=magenta,
	filecolor=cyan,
	linkbordercolor={1 0 0},
	citebordercolor={0 1 0},
	urlbordercolor={0 1 1},
} 
\usepackage{cleveref}
\usepackage[mathlines]{lineno}
\usepackage{float,wrapfig}
\begin{document}
	\preprint{APS/123-QED}
	
	\title{Interplay of energy, dissipation, and error in kinetic proofreading: Control via concentration and binding energy}
	\author{Premashis Kumar}
	\affiliation{S. N. Bose National Centre For Basic Sciences, Block-JD, Sector-III, Salt Lake, Kolkata 700 106, India}
	\author{Kinshuk Banerjee}
	\affiliation{Department of Chemistry, Acharya Jagadish Chandra Bose College, Kolkata 700 020, India}
	\author{Gautam Gangopadhyay}
	\email{gautam@bose.res.in}
	\affiliation{S. N. Bose National Centre For Basic Sciences, Block-JD, Sector-III, Salt Lake, Kolkata 700 106, India}
	\date{\today}
	
	\begin{abstract}
Kinetic proofreading mechanisms explain the extraordinary accuracy observed in central biological events in terms of the enhanced specificity of substrate selection networks under a nonequilibrium environment. The nonequilibrium steady state theory incorporated with a chemical thermodynamic framework is implemented to execute a systematic investigation of dynamic and thermodynamic features of the proofreading network under continuous fuel consumption. We have identified that the dissipation-error trade-off domain of the network has a one-to-one correspondence with the deeper portion of the basin-like error rate profile depicted here. Further, quantifying the energy cost through concentration control of the chemical fuel aids in unveiling the association of the energy and chemical work with the optimal operating region of the biological error-correcting mechanism. It is shown that the proper energy content of the system, the semigrand Gibbs free energy, approaches a nominal value in the trade-off regime, whereas it gets considerably minimized in the domain with a lack of trade-off. We have also introduced a performance measuring entity corresponding to different energetic discrimination magnitudes as a product of average dissipation and coefficient of variation for the whole range of chemical fuel. Besides invoking a different perspective to the experimental and theoretical assessment of biological processes like DNA replication or protein synthesis, our findings can be advantageous in designing more efficient synthetic biological architectures. 
	\end{abstract}
	\maketitle
	\section{\label{sec:1} Introduction}
	Biosynthesis, a complex process comprising a series of chemical reactions, supplies essential composite products required for various necessary processes related to the survival, regulation, and growth of living organisms with extraordinary accuracy. The observation of remarkably high accuracy in biosynthesis processes is rationalized by the strategy known as kinetic proofreading proposed independently by J. J. Hopfield and J. Ninio~\citep{Hopfield4135, NINIO1975587}. The proofreading mechanism asserts that the high fidelity in selecting the right substrate over wrong ones can be adopted in the reaction network model of core biological events like DNA replication~\citep{johnson1993, SCHAAPER199323762}, RNA transcription~\citep{Blank, Goute} or protein synthesis~\citep{ZAHER2009746} by strongly driving the network out of equilibrium. Most importantly, biological systems utilize the chemical energy of energy-bearing biomolecules to drive the process. Therefore, in the simplest form, the kinetic proofreading mechanism of the biosynthesis process can be cast into the cyclic receptor-ligand binding model~\citep{Qian1, Qian2} coupled with Adenosine Tri-Phosphate(ATP) hydrolysis. Interestingly, a reaction network driven out of equilibrium approaches a nonequilibrium steady state(NESS) characterized by the nonzero driving force and reaction flux which generates dissipation~\citep{qiandissi} in the system. Dissipation plays a crucial role in sustaining or enhancing the performance of a reaction network. Particularly, in the context of kinetic proofreading network, a trade-off between the dissipation and error rate can exist~\citep{BENNETT1979, Murugan12034, Mallory} and a proper investigation of this trade-off scenario aids to detect optimal operating regimes of the network and biological systems having such network. Nevertheless, the lowest error rate can be achieved for vanishingly small dissipation for the energetic discrimination scenario implying the difference of binding energy among intermediates~\citep{Hopfield4135, NINIO1975587} of the network. A generalized model~\citep{murugan2012, Murugan12034, wong2018} of error-correcting mechanism also exploited the energetic discrimination scenario to investigate substantial complex and essential features of the proofreading scheme, including error-dissipation relation. 
	
	The error-dissipation constraint of the proofreading scheme has been studied in different variants of proofreading networks with the context of relevant biological processes to reveal the guiding role of dissipation~\citep{kbgg, Mallory}. In this regard, the overall energy cost of the mechanism plays a pivotal role in shaping and carrying out these biological process~\citep{Savageau1979EnergyCO, Mehta17978}. Thus the energetic expenditure of detecting the right substrate would impose an important constraint on the proper functioning of the network given the limited energy budget in a biological environment. Although much attention has been dedicated to understanding the interplay between error and dissipation, the explicit energy utilization and modulation in a chemical fuel-driven error-correcting mechanism is not addressed properly. However, the recent development in nonequilibrium thermodynamics of open chemical reaction network(CRN)~\citep{Rao2016NonequilibriumThermodynamics, Falasco2018InformationPatterns} provides a rigorous basis to quantify the proper energy content of an arbitrarily driven system under the concentration and flux control~\citep{Avanzini2021ThermodynamicsOC} of the exchanged species. Such energy quantification concerning externally-controlled chemical fuel in the context of proofreading schemes can provide crucial insight into the concept of chemical work and the thermodynamic efficiency of biological information processing mechanisms. Additionally, as the existence of multiple regimes of different discriminatory ability has been reported~\citep{murugan2012} recently, it will be very interesting to explore how the system's energy under a nonequilibrium environment is assigned to different discriminatory regimes having various possible trade-off situations. The resemblance of different proofreading regimes with stages of microtubule growth~\citep{Murugan12034} and nonequilibrium sensing~\citep{Hartich_2015} seeks a complete thermodynamic picture of such state to unveil the underlying general thermodynamic principles applicable to biological information processing~\citep{Lan2012TheET}.  
	
	Here, we aim to unveil the overall performance of the kinetic proofreading network in terms of error, dissipation, and proper energy representation of the open system and establish a connection among these entities within different discriminatory regimes. In this regard, our work further intends to shed light on accessing the different error rates and associated dissipation by merely controlling the concentration of certain species and magnitude of the energetic discrimination without changing the network structurally and then evaluating the performance of the proofreading network. We have considered a simple two-cycle network of kinetic proofreading to systematically employ recently developed conservation law~\citep{Polettini2014IrreversibleLaws, conlawPolettini} based analytically tractable chemical thermodynamic framework~\citep{Rao2016NonequilibriumThermodynamics, Falasco2018InformationPatterns} to quantify the network energetics properly. Hence, along with the conventional nonequilibrium Gibbs free energy, we would evoke the semigrand Gibbs free energy~\citep{Rao2016NonequilibriumThermodynamics, Falasco2018InformationPatterns} to identify the role of system energetics in the performance of the network. In this regard, measuring energy and dissipation by steering the concentrations of certain chemical fuels like ATP would be more natural in a driven system like this. Moreover, we will introduce an overall performance measuring quantity that can assess the basis of overall error-dissipation trade-off scenarios for different energetic discrimination magnitudes. Besides theoretical importance, our approach of manipulating error, dissipation, and energy with varying concentrations of energy-rich species can indeed open up various implementation opportunities in synthetic reaction networks and biological systems.  
	
	The paper is organized as follows. First, we have described the kinetic network of the proofreading mechanism and corresponding NESS dynamics in sec.\ref{sec:II}. The error rate of the network as a function of externally controllable parameters and the binding energy difference is also discussed in the section. In the next section, we have captured the thermodynamic response of the network due to ATP concentration variation. We have also formulated the semigrand Gibbs free energy by exploiting the network topology. Then, the error-dissipation-semigrand Gibbs free energy profile and quantification of the network performance are depicted in sec.\ref{sec:VI}. Finally, the summary and conclusions are provided.
	\section{\label{sec:II}Kinetic Network: Nonequilibrium Steady State Characterization and Error rate Variation}
	\begin{figure*}
		\includegraphics[width=\textwidth]{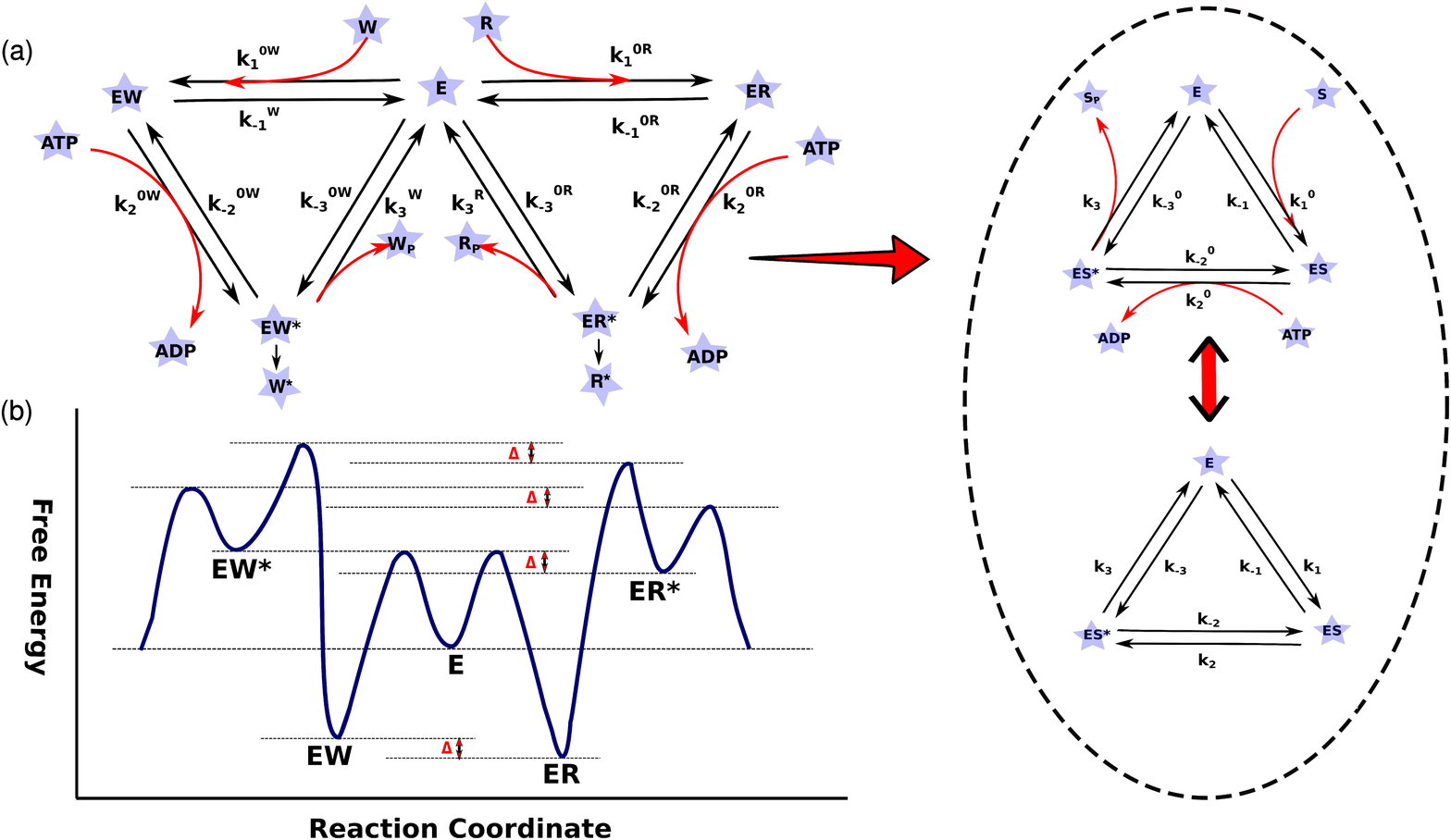}
		\caption {\label{fig:proofreadingmodel}kinetic model: In \cref{fig:proofreadingmodel}(a), two distinct cycles correspond to two very similar substrates `right'(`R') and `wrong'(`W'). `R\textsuperscript{*}' and `W\textsuperscript{*}' are related to the synthesis rate, which is assumed to be sufficiently low to be neglected here. Species `W', `R', `ATP', `ADP', `R\textsubscript{p}' and `W\textsubscript{p}' are assumed to have constant concentrations over the time scale of interest and can be absorbed within rate constants.  In \cref{fig:proofreadingmodel}(b), the hypothetical image of the free energy landscape for the kinetic network has been considered here. Right: Any cycle of the kinetic proofreading network can be represented by a single cyclic model shown in the upper panel. By exploiting pseudo-first-order rate constants  $k_{1}=k_{1}^0(S), k_{-3}=k_{-3}^0(S_p), k_{+2}=k_{+2}^0(ATP)$ and $k_{-2}=k_{-2}^0(ADP)$, the single cyclic model in the upper panel can be effectively mapped into a simple three-state system as shown in the lower panel, with seemingly no explicit material exchange with the environment.}
	\end{figure*}
	The proofreading network is described here with a nominal synthesis rate. Indeed, in the presence of energetic discrimination only, Hopfield's model~\citep{Hopfield4135} shows minimal error at zero velocity. A schematic diagram of the kinetic network of the proofreading scheme is illustrated in \cref{fig:proofreadingmodel}(a). One cycle of the double-cycle proofreading model corresponds to the `right'(`R') pathway, whereas another cycle belongs to the `wrong'(`W') one. The network depicts that a protein E binds with substrate `R' (or `W') to form intermediate complexes `ER', `ER\textsuperscript{*}' (or `EW', `EW\textsuperscript{*}')  and then disassociates into enzyme and product. The discrimination between `R' and `W' substrates in a living system usually occurs through unbinding rather than binding~\citep{McKeithan5042}. Hence, it is comprehensible to consider that all rate constants, except those corresponding to unbinding, are equal for `R' and `W' pathways because of the structural similarity of these two substrates. Thus, it is sufficient to assume that difference in affinity due to only rate constants $k_{-1}^R \neq k_{-1}^W$ and $k_{3}^R \neq k_{3}^W$ in \cref{fig:proofreadingmodel}(a). Furthermore, with correct substrate(`R') having more residence time, $k_{-1}^{R}$ and $k_{3}^{R}$ are connected with the corresponding rates of `W' reaction cycle through the relation $\frac{k_{-1}^{R}}{k_{-1}^{W}}=\frac{k_{3}^{R}}{k_{3}^{W}}=\exp^{-\Delta}=\sigma <1$ where $\Delta$ is the binding energy difference between `W' and `R' intermediate complexes. Now depending on the concentration dynamics of the species, we can divide all the species of the proofreading network in \cref{fig:proofreadingmodel}(a) into two disjoint sets of `I' and `C',
	$$\frac{\{E, ER, ER^*, EW, EW^*\}}{I}\bigcup\frac{\{R, ATP, ADP, R_p, W, W_p\}}{C},$$
	where `I' is a set of intermediate species having dynamic concentration, and `C' is a set of chemostatted species with a constant concentration within the time scale of interest. In other words, elements of the set `I' form the vertices in the graphical representation of a reaction network, whereas elements of set `C' are associated with the edge. To keep the approach more general, $R_p$ and $W_p$ instead of commonly used $R$ and $W$ have been considered in the third step of the proofreading network. 
	
	Now the schematic diagram, \cref{fig:proofreadingmodel}(a) yields $eR^*+eR+eW+eW^*+e=e_0$ with $e_0$ being the total enzyme concentration and $eR, eR^*, eW, eW^*$ being concentrations of species `ER', `ER\textsuperscript{*}', `EW', `EW\textsuperscript{*}', respectively. We assume that total enzyme concentration can be represented as $e_0=e_0^{R}+e_0^{W}$ to simplify the problem considerably and generally $e_0^{R} \neq e_0^{W}$. This assumption results in $eR^*+eR+e^R=e_0^{R}$ and  $e^W+eW+eW^*=e_0^{W}$ where $e=e^R+e^W$. With these considerations, we can mathematically map this two cycle proofreading model into a single cycle model as shown in right panel of \cref{fig:proofreadingmodel}. Hence kinetics of the cycle involving right substrate can be expressed as
	\begin{subequations}
		\begin{align}
			\frac{\partial eR}{\partial t}&=k_{1}^R(e_0^R-eR^*-eR)-(k_{-1}^R+k_{+2}^R)(eR)+k_{-2}^R(eR^*)\label{d3}\\
			\frac{\partial eR^*}{\partial t}&=k_{-3}^R(e_0^R-eR^*-eR)-(k_{-2}^R+k_{+3}^R)(eR^*)+k_{+2}^R(eR).
			\label{d4}
		\end{align}
	\end{subequations}   
	Here  $k_{1}=k_{1}^0(R), k_{-3}=k_{-3}^0(R_p), k_{+2}=k_{+2}^0(ATP)$ and $k_{-2}=k_{-2}^0(ADP)$ are pseudo-first-order rate constants~\citep{Qian1, Qian2} and concentration of ATP, ADP, R, and $\mathrm{R_{p}}$ are held at constant levels. Exploiting the condition $eR^*+eR+e^R=e_0^{R}$, we acquire following steady state concentrations from eq. \eqref{d3} and \eqref{d4},
	\begin{widetext}
		\begin{subequations}
			\begin{align}
				eR_{ss}=\frac{(k_{-2}^{R}k_{-3}^{R}+k_1^{R}k_3^R+k_1^{R}k_{-2}^{R})e_0^{R}}{(k_1^{R}k_{-2}^{R}+k_1^{R}k_3^{R}+k_1^{R}k_2^{R})+(k_{-1}^{R}k_{-3}^{R}+k_2^{R}k_{-3}^{R}+k_{-2}^{R}k_{-3}^{R})+(k_{-1}^{R}k_{-2}^{R}+k_{-1}^{R}k_3^{R}+k_2^{R}k_3^{R})} \label{s3}\\
				eR^*_{ss}=\frac{(k_1^Rk_2^R+k_{-1}^Rk_{-3}^R+k_2^Rk_{-3}^R)e_0^R}{(k_1^Rk_{-2}^R+k_1^Rk_3^R+k_1^Rk_2^R)+(k_{-1}^Rk_{-3}^R+k_2^Rk_{-3}^R+k_{-2}^Rk_{-3}^R)+(k_{-1}^Rk_{-2}^R+k_{-1}^Rk_3^R+k_2^Rk_3^R)}\label{s5}\\
				e_{ss}^R=\frac{(k_{-1}^Rk_{-2}^R+k_{-1}^Rk_3^R+k_2^Rk_3^R)e_0^R}{(k_1^Rk_{-2}^R+k_1^Rk_3^R+k_1^Rk_2^R)+(k_{-1}^Rk_{-3}^R+k_2^Rk_{-3}^R+k_{-2}^Rk_{-3}^R)+(k_{-1}^Rk_{-2}^R+k_{-1}^Rk_3^R+k_2^Rk_3^R)}.\label{s6}
			\end{align}
		\end{subequations}
	\end{widetext}
	Now, unlike a equilibrium steady state, there will be a clockwise steady state flux for the cyclic reaction network at NESS. Each cycle of the proofreading network comprises of three elementary reactions and  net fluxes of these reactions elementary can be obtained as
	$j_1=k_1(e)-k_{-1}(es);j_2=k_2(es)-k_{-1}(es^*) \text{and}  j_3=k_3(es^*)-k_{-3}(e)$. Therefore, the steady state cycle flux in clock-wise direction for a single cycle is represented as 
	\begin{widetext}
		\begin{equation}
			j_{ss}^R=\frac{(k_1^Rk_2^Rk_3^R-k_{-1}^Rk_{-2}^Rk_{-3}^R)e_0^R}{(k_1^Rk_{-2}^R+k_1^Rk_3^R+k_1^Rk_2^R)+(k_{-1}^Rk_{-3}^R+k_2^Rk_{-3}^R+k_{-2}^Rk_{-3}^R)+(k_{-1}^Rk_{-2}^R+k_{-1}^Rk_3^R+k_2^Rk_3^R)}.
			\label{f2}
		\end{equation} 
	\end{widetext}
	In the case of wrong substrate pathway, $eW_{ss}, eW_{ss}^*, e_{ss}^W$ and $j_{ss}^W$ will have the same expression with all the superscript `W' instead of `R' in each term. As required by a thermodynamic viewpoint, all the reaction steps have to be reversible in nature. The driving force associated with the NESS cycle flux is the net chemical potential around the single cycle,  
	$\mu_L=\mu_{e,es}+\mu_{es,es^*}+\mu_{es^*,e}=\ln{\frac{k_1k_2k_3}{k_{-1}k_{-2}k_{-3}}} =\ln{\gamma}.$ As the rate constants $k_1, k_{\pm 2}, k_{-3}$ are equal in both `W' and `R' pathways and $k_{-1}$ and $k_{3}$  are connected through a common factor $\exp^{-\Delta}$, both the `W' and `R' cycles are subjected to equal driving force. We have set Boltzmann's constant $k_B=1$ and temperature $T=1$ to slim our notations throughout this study. Unless otherwise stated, we have implemented following rate constants throughout our study, $k_1^R=k_1^W=5$, $k_3^R=1$, $k_{-1}^R=50$, $k_{-2}^R=k_{-2}^W=10.^{-3}$ and $k_{-3}^R=k_{-2}^W=10.^{-3}$~\citep{Banerjee2017Accuracy}. Rate constants with equal magnitudes for `R' and `W' cycle have been used without a superscript in future discussion.  
	
	\subsection{Error Rate Profile}\label{error_basin}
	\begin{figure*}
		\begin{center}
			\includegraphics[width=\textwidth]{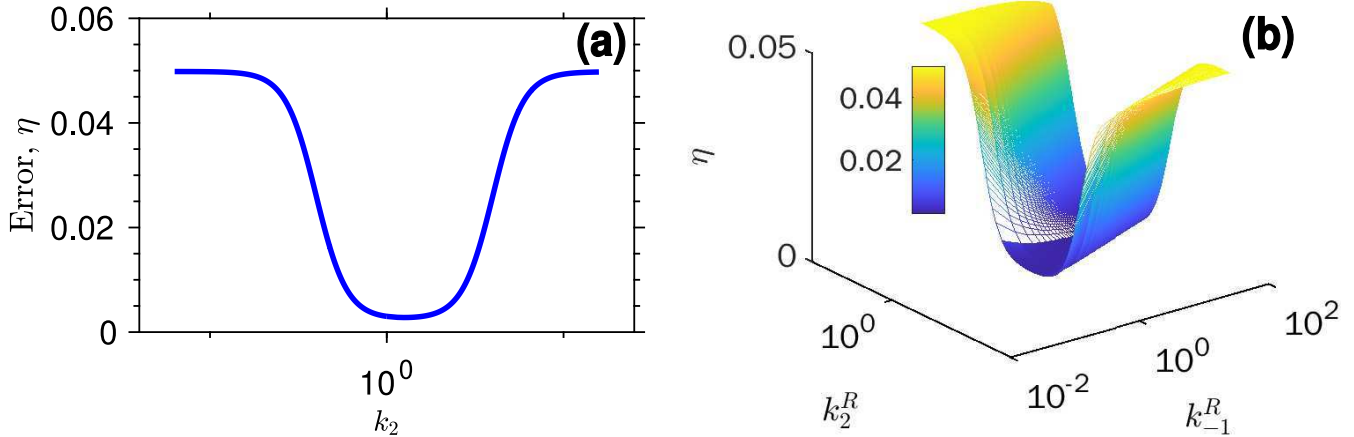}
			\caption{\label{fig:1} In \cref{fig:1}(a), a basin-like characteristic of error rate is obtained by varying only $k_2$ from $10^{-6}$ to $10^{6}$. Other rate constants(or pseudo-first-order rate constants) of the proofreading network are fixed at $k_1=5$, $k_3^R=1$, $k_{-1}^R=50$, $k_{-2}=10^{-3}$ and $k_{-3}=10^{-3}$. To visualize the characteristics of the error rate more lucidly, we have taken the log scale of the $k_{+2}$. In \cref{fig:1}(b), a 3D counterpart of error basin is illustrated by changing both $k_{-1}$ and $k_{+2}$ simultaneously. $k_{+2}$ is varied from $10^{-5}$ to $10^{5}$, while $k_{-1}$ is changed within the range, $0.5-50$. Other rate constants are the same as in \cref{fig:1}(a). The binding energy difference, $\Delta=3$ in both figures.}
		\end{center}
	\end{figure*}
	Initially we would examine the error rate profile by varying multiple pseudo-first-order rate constants simultaneously. We have expressed the error rate of the proofreading scheme by mapping the two-cycle model into a single cycle. In the limit of slow synthesis rate, the error rate of the kinetic proofreading network at NESS can be defined as~\citep{murugan2012}
	\begin{equation}
		\eta =\frac{eW^*_{ss}}{e_{ss}^W}\frac{e_{ss}^R}{eR^*_{ss}} \label{error}
	\end{equation}
	where $eW^*_{ss}$ and $eR^*_{ss}$ are steady-state concentrations of `EW\textsuperscript{*}' and `ER\textsuperscript{*}' respectively and concentration of enzyme corresponding `R' and `W' cycles is given by $e_{ss}^R$ and $e_{ss}^W$. Under equilibrium condition, the error rate of the kinetic proofreading network  would result in $\eta=\exp^{-\Delta}$. However, an error rate as low as $\eta=\exp^{-2\Delta}$ can be achieved  within a nonequilibrium environment. 
	
	Now using eq. \eqref{s5} and \eqref{s6}, we can express the error rate
	from eq. \eqref{error} as,
	\begin{equation}
		\eta=\frac{(k_{1}^{W}k_{2}^{W}+k_{-1}^{W}k_{-3}^W+k_{2}^{W}k_{-3}^{W})(k_{-1}^Rk_{-2}^R+k_{-1}^Rk_3^R+k_2^Rk_3^R)}{(k_{1}^{R}k_{2}^{R}+k_{-1}^{R}k_{-3}^R+k_{2}^{R}k_{-3}^{R})(k_{-1}^Wk_{-2}^W+k_{-1}^Wk_3^W+k_2^Wk_3^W)}.\label{generalerror}
	\end{equation}
	Natural dominance of reaction step 2 rather than step 3 of the proofreading network in production of substrate-enzyme complex $eR^*$ or $eW^*$ implies $k_1>>k_{-3}$ in \cref{fig:proofreadingmodel}. Further, preference of step 3  in the formation  of `R\textsubscript{p}' (or `W\textsubscript{p}') from  $eR^*$ (or $eW^*$) suggests $k_3>>k_{-2}$. Taking very small values of $k_{-3}$ and $k_{-2}$ ensure that corresponding steps of reaction network are almost irreversible. Now under assumptions $k_1>>k_{-3}$, and $k_3>>k_{-2}$, we would analyze the modification of the specificity of proofreading network due to the relative strength between rate constants $k_{-1}$ and $k_{+2}$. For this purpose, we have taken three different situations on the basis of relative strength between rate constants $k_{-1}$ and $k_{+2}$ and identified orders of the specificity in those situations.
	\begin{itemize}
		\item   $k_{-1} >> k_{+2}$; $\gamma=\frac{k_1^Rk_2^Rk_3^R}{k_{-1}^Rk_{-2}^Rk_{-3}^R}\approx 1$: The general expression of the error rate in eq. \eqref{generalerror} becomes
		$\eta_1=\frac{k_{-1}^{W}}{k_{-1}^{R}}\frac{(k_{-1}^{R}k_{3}^R)}{(k_{-1}^{W}k_{3}^{W})}=\exp^{-\Delta}=\sigma.$\end{itemize}
	\begin{itemize}
		\item   $k_{-1} \simeq k_{+2}$; $\gamma>>1$: The error rate  in eq. \eqref{generalerror} results in  $\eta_2=\frac{(k_{-1}^Rk_3^R)}{(k_{-1}^Wk_3^W)}\frac{(1+\frac{k_2^R}{k_{-1}^R})}{(1+\frac{k_2^W}{k_{-1}^W})}=\sigma^2\frac{(1+\frac{k_2^R}{k_{-1}^R})}{(1+\frac{k_2^W}{k_{-1}^W})}=\sigma^2.$ 
	\end{itemize}
	\begin{itemize}
		\item   $k_{-1} << k_{+2}$; $\gamma>>1$: Finally this condition yields error rate as, $\eta_3=\frac{k_3^R}{k_3^W}=\sigma.$
	\end{itemize}
	This result reveals the subtle role of the relative strength of rate constants $k_{-1}$ and $k_{+2}$ in the modification of error. We can use either $k_2$ or $k_{-1}$ as a control parameter of this proofreading scheme to access different discriminatory regimes. Most importantly, for the condition $k_{-1} << k_{+2}$, although the system is far from equilibrium, we do not obtain any improvement over the error rate near equilibrium. 
	
	We have obtained an error rate profile in \cref{fig:1}(a) from the general expression of the error rate in eq. \eqref{generalerror} by varying only rate constant $k_{+2}$ at fixed values of binding energy difference $\Delta$ and all other rate constants. As we proceed from $k_{-1} >> k_{+2}$ to $k_{-1} \simeq k_{+2}$ regime by varying $k_{+2}$, it is evident from \cref{fig:1}(a) that there is an abrupt fall in the error and eventually the error will reach its minimum value $\sigma^2$. The plateau region in \cref{fig:1}(a) represents the $k_{-1} \simeq k_{+2}$ regime $\sigma^2$. Then, while we move in parameter space from the $k_{-1} \simeq k_{+2}$ region to $k_{-1} << k_{+2}$, a basin-like characteristics of the error emerges as the error rate approaches the equilibrium error rate value $\sigma$ within nonequilibrium environment. This graphical representation of the general error rate in \cref{fig:1}(a) depicts the analytical results under three conditions, as discussed earlier in this section. In \cref{fig:1}(b), we have obtained a 3D counterpart of the error basin in \cref{fig:1}(a) by varying both $k_{-1}$ and $k_{+2}$ simultaneously. This 3D figure more clearly illustrates the modulation of error rate due to relative strength between $k_{-1}$ and $k_{+2}$. Thus, this representation will be crucial to ensure maximum accuracy of the proofreading scheme by controlling either $k_{-1}$ or $k_{2}$ or both rate constants at the same time for a fixed $\Delta$($=3$ here). 
	
	The central observation from this analysis is that error rate $\eta$ is not constant under the nonequilibrium condition but strongly depends on the relative strength between rate constants $k_{-1}$ and $k_{+2}$. We assert that the kinetic proofreading network can have two different regimes with distinct discrimination ability depending on the relative strength of rate constants $k_{-1}$ and $k_{+2}$  under nonequilibrium conditions. As the pseudo-first-order rate constants contain the concentration of chemostatted species by definition, we can also modify the error rate curve in the same way by controlling the concentration of these species. 
	
	\subsection{\label{deltaim}Impact of Binding Energy Difference}    
	\begin{figure*}
		\begin{center}
			\includegraphics[width=\textwidth]{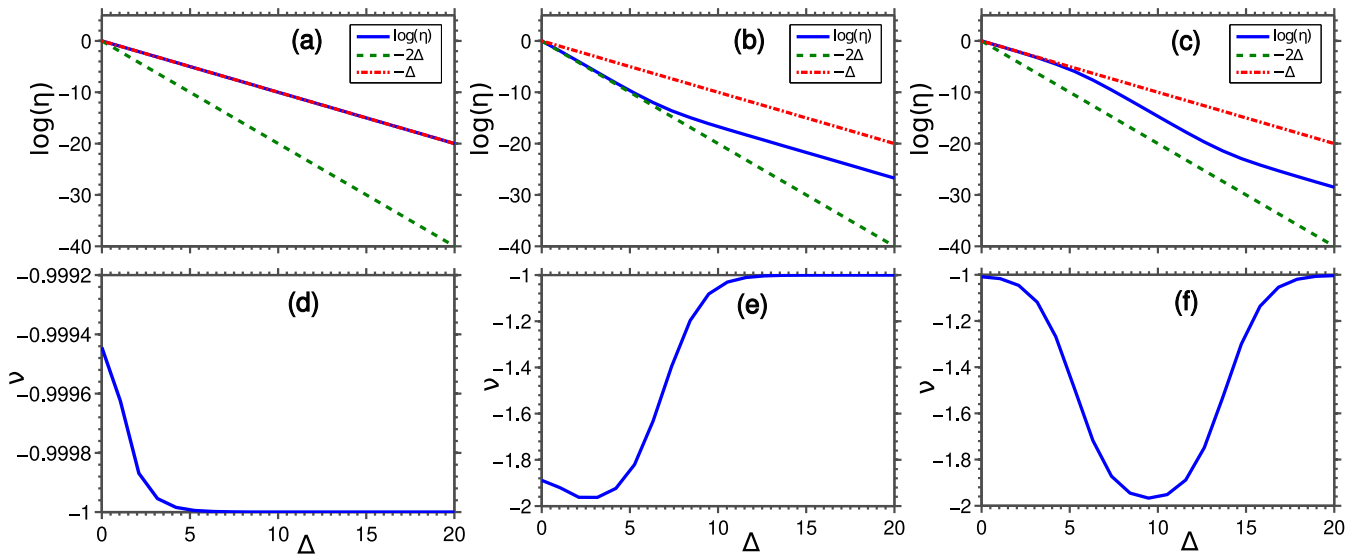}
			\caption{\label{fig:3} The impact of the binding energy difference, $\Delta$ on the error rate, $\eta$ for three different regimes:  $\log{\eta}$ vs $\Delta$ is exhibited here for three regions corresponding to  $k_{-1} >> k_{+2}$, $k_{-1} \simeq k_{+2}$ and $k_{-1} << k_{+2}$ in \cref{fig:3}(a), \cref{fig:3}(b) and \cref{fig:3}(c), respectively. The corresponding variations of the discriminatory index, $\nu=\frac{\partial \log(\eta)}{\partial \Delta}$ over the range of $\Delta$ for $k_{-1} >> k_{+2}$, $k_{-1} \simeq k_{+2}$ and $k_{-1} << k_{+2}$ are depicted in \cref{fig:3}(d), \cref{fig:3}(e) and \cref{fig:3}(f), respectively.}
		\end{center}
	\end{figure*}
	To capture the impact of binding energy difference, $\Delta$  on the error rate of the kinetic proofreading network, we have shown variation of  $\log(\eta)$ with respect to $\Delta$, separately for three different regimes corresponding to the relative strength between $k_{-1}$ and $k_{+2}$. Then akin to the discriminatory index in ref.~\citep{murugan2012}, we have defined an index, $\nu=\frac{\partial \log(\eta)}{\partial \Delta}$, i.e., the slope of the $\log(\eta)$ versus $\Delta$ plot. For example, this definition yields $\nu=-1$ for the equilibrium condition as $\eta=\exp^{-\Delta}$ in this regime.  
	
	Now we would analyze the error rate, $\eta$ in terms of the newly defined quantity, $\nu$, for three previously mentioned regimes. For the regime corresponding to $k_{-1} >> k_{+2}$, $\log(\eta)$ follows line corresponding to $\exp{-\Delta}$ over the whole range of $\Delta$ as shown in \cref{fig:3}(a). As all other parameters are kept fixed at constant values, we can assert that the error rate in this  regime is not effected by $\Delta$. The same is evident from \cref{fig:3}(d) as $\nu=-1$ for the whole range of $\Delta$. For the regime, $k_{-1} \simeq k_{+2}$, $\log(\eta)$ curve in \cref{fig:3}(b) fairly follows the line corresponding to $\exp{-2\Delta}$ up to a certain extent of $\Delta$ magnitude and then it deviates from this line upon further increment of $\Delta$. It is clear from the \cref{fig:3}(e) that $\nu$ in this regime switches from $-2$ to $-1$ for higher $\Delta$ value. Hence, the accuracy is less than its maximum level for the parameter region representing the relatively higher binding energy difference. The error rate in such case can be generally represented by $\eta=\exp{-(1+\beta)\Delta}$ with $0<\beta<1$. Finally, for the regime, $k_{-1} << k_{+2}$, $\log(\eta)$ at first follows the line of $\exp{-\Delta}$ in \cref{fig:3}(c) as expected from the error basin in \cref{fig:1}(a). However, as the $\Delta$ is further extended over a certain value, the  $\log(\eta)$ line gradually deviates from the $\exp{-\Delta}$ line to the $\exp{-2\Delta}$. This deviation for higher values of $\Delta$ suggests that it is possible to have a relatively more effective proofreading(with respect to the lower $\Delta$ counterpart) even for regime corresponding to $k_{-1} << k_{+2}$. The error rate curve again moves towards the $\exp{-\Delta}$ line on further increment of $\Delta$  above a certain value. The whole observation in \cref{fig:3}(c) is also confirmed by the variation of the $\nu$ between $-1$ and $-2$ in \cref{fig:3}(f). Thus, the overall graphical analysis in \cref{fig:3} reveals that it is possible to expand, shrink, or shift the parameter space corresponding to a particular error rate by using binding energy difference as a control parameter. 
	
	\section{Thermodynamic Response Due to Concentration Control} 
	Any change in concentrations of chemostatted species can be a source of massive modification in the system's entropic and energetic response under a nonequilibrium environment~\citep{pkgg,pkgg2}. The pseudo-first-order rate constants are defined in terms of concentrations of those chemostatted species present in the open chemical reaction network representation of the kinetic proofreading. Therefore, rather than changing the rate constants, the variation in the pseudo-first-order rate constants can be realized by altering the chemostatted concentrations. In the following study, we will analyze the thermodynamic evolution of the proofreading network controlled by the influx of the chemostatted species, especially the fuel molecule ATP, and establish the association of this thermodynamic response with the dynamics of the network.     
	\subsection{Conservation Laws and Emergent Cycles}
	The reaction network of the kinetic proofreading in \cref{fig:proofreadingmodel}(a) comprises of six elementary chemical reactions and eleven species. Hence, the proofreading network can be presented by the following $11\times 6$ stoichiometric matrix~\citep{Alberty2003ThermodynamicsReactions} 
	\begin{gather}
		S_{\rho}^{\sigma}=
		\bordermatrix{ ~ & R_{1} & R_{2}&R_{3}&R_{4}&R_{5}&R_{6}\cr
			E&-1 &0&1&-1&0&1\cr
			ER&1&-1&0&0&0&0 \cr
			ER^*&0&1&-1&0&0&0 \cr
			EW&0&0&0&1&-1&0\cr
			EW^*&0&0&0&0&1&-1 \cr
			R&-1&0&0&0&0&0\cr
			ATP&0&-1&0&0&-1&0\cr
			ADP&0&1&0&0&1&0\cr
			W&0&0&0&-1&0&0\cr
			R_p&0&0&1&0&0&0\cr
			W_p&0&0&0&0&0&1\cr}. \label{st}
	\end{gather} 
	The left null linear independent vectors corresponding to left null space of this stoichiometric matrix $S_{\rho}^{\sigma}$ are known as the conservation laws~\citep{Alberty2003ThermodynamicsReactions} and can be expressed as, 
	$\sum_{\sigma}{l_{\sigma}^{\lambda}S_{\rho}^{\sigma}}=0 \label{cons}$ where $\{l_{\sigma}^{\lambda}\}\in \mathbb{R}^{(\sigma-w )\times \sigma}, w=rank(S_{\rho}^{\sigma}).$
	For the stoichiometric matrix \eqref{st} of the proofreading network, the following five vectors represent the conservation laws of the closed reaction network,
	\begin{widetext}
		\begin{gather}
			l_{\sigma}^{\lambda=1}=  
			\bordermatrix{~&E&ER&ER^*&EW&EW^*&R&ATP&ADP&W&R_p&W_p\cr
				&1&1&1&1&1&0&0&0&0&0&0\cr}\nonumber\\
			l_{\sigma}^{\lambda=2}=  
			\bordermatrix{~&E&ER&ER^*&EW&EW^*&R&ATP&ADP&W&R_p&W_p\cr
				&0&0&0&0&0&0&1&1&0&0&0\cr}\nonumber\\
			l_{\sigma}^{\lambda=3}=  
			\bordermatrix{~&E&ER&ER^*&EW&EW^*&R&ATP&ADP&W&R_p&W_p\cr&0&1&0&1&0&1&-1&0&1&0&0\cr}\nonumber\\
			l_{\sigma}^{\lambda=4}=  
			\bordermatrix{~&E&ER&ER^*&EW&EW^*&R&ATP&ADP&W&R_p&W_p\cr&0&1&1&0&0&1&0&0&0&1&0\cr}\nonumber
		\end{gather}
		and
		\begin{gather}  
			l_{\sigma}^{\lambda=5}=  
			\bordermatrix{~&E&ER&ER^*&EW&EW^*&R&ATP&ADP&W&R_p&W_p\cr
				&-1&-2&-1&-1&0&-1&1&0&0&0&1\cr}.\nonumber
		\end{gather}
	\end{widetext}
	These conservation laws of the reaction network further specify conserved quantities of the network known as Components of the system and are defined as, 
	$ L_{\lambda}=\sum_{\sigma}{l_{\sigma}^{\lambda}}z_{\sigma}$ such that $\frac{d}{dt}\int dr L_{\lambda}=0$.
	Thus, components corresponding to those five conservation laws are $L_1=e+eR+eR^*+eW+eW^*;$ $L_2=ATP+ADP;$
	$L_3=eR+eW+R-ATP+W;$
	$L_4=eR+eR^*+R+R_p$
	and $L_5=-e-2eR-eR^*-eW-R+ATP+W_p$. 
	
	While a closed system is opened by chemostatting, the stoichiometric matrix $S_{\rho}^{\sigma}$ breaks into two parts: $S_{\rho}^{I}$ referring to intermediate species and $S_{\rho}^{C}$ belonging to chemostatted species. Therefore, conservation laws in an open system are characterized as
	\begin{equation}
		l_{I}^{\lambda}S_{\rho}^{I}+l_{C}^{\lambda}S_{\rho}^{C}=0
		\begin{cases}
			l_{I}^{\lambda_{b}}S_{\rho}^{I}\neq 0 &\text{ broken CL},\\
			l_{I}^{\lambda_{u}}S_{\rho}^{I}= 0&\text{ unbroken CL}
			\label{condi}
		\end{cases}
	\end{equation}
	with $u$ and $b$ being labels corresponding to unbroken and broken ones, respectively and $\{l^{\lambda}\}=\{l^{\lambda_b}\}\cup\{l^{\lambda_u}\}$. In this reaction network, conservations laws, $l_{\sigma}^{\lambda=2}, l_{\sigma}^{\lambda=3}, l_{\sigma}^{\lambda=4}$ and $l_{\sigma}^{\lambda=5}$ of the closed system can be broken due to chemostatting. Depending on whether chemostatted species break a conservation law or not, the set of chemostatted species can be divided into two subsets $\{C\}=\{C_{b}\}\cup\{C_{u}\}$. It is apparent from eq. \eqref{condi} that the broken conservation laws are not left null vectors of $S_{\rho}^{I}$ for at least one reaction of the network. Hence, components of the open system related to the broken conservation laws are no longer a global conserved quantities and are denoted as $L_{\lambda_b}$. The unbroken conservation law of the open system of the proofreading are obtained from the stoichiometric matrix of intermediate species, $S_{\rho}^{I}$ as, 
	\begin{gather}  
		l_{I}^{\lambda_u=1}=  
		\bordermatrix{~&E&ER&ER^*&EW&EW^*\cr
			&1&1&1&1&1\cr}\nonumber
	\end{gather}
	and corresponding component $L_{\lambda_u}=e+eR+eR^*+eW+eW^*$ refers to the total concentration of the enzyme, a global conserved quantity of this open system. Right null vectors of the stoichiometric matrix, $S_{\rho}^{\sigma}c_{\sigma}^n$ represent internal cycles which restore all the species' concentration to its prior state upon completion. However, this kinetic proofreading network has no internal cycle. Instead, this open reaction network contains following two independent emergent cycles~\citep{Polettini2014IrreversibleLaws}, 
	\begin{gather}  
		c_{1}^{}=  
		\bordermatrix{~\cr
			1&1\cr
			2&1\cr
			3&1\cr
			4&0\cr
			5&0\cr
			6&0}\nonumber
		\hspace{10 mm}
		\text{and}
		\hspace{10 mm}
		c_{2}^{}=  
		\bordermatrix{~\cr
			1&0\cr
			2&0\cr
			3&0\cr
			4&1\cr
			5&1\cr
			6&1}.
	\end{gather}  
	A complete emergent cycle keeps the states of intermediate species unchanged but chmeostatted species are exchanged between system and chemostats~\citep{Rao2016NonequilibriumThermodynamics}. The net stoichiometry of the emergent cycle  $c_1$ and $c_2$ are \ch{R +  ATP <-> R_p + ADP} and \ch{W +  ATP <-> W_p + ADP}, respectively. The steady-state current vectors for the emergent cycle $c_1$ and $c_2$ are
	\begin{gather}  
		j_{c_{1}}^{}=  
		\bordermatrix{~\cr
			1&j_{c_1}\cr
			2&j_{c_1}\cr
			3&j_{c_1}\cr
			4&0\cr
			5&0\cr
			6&0}\nonumber
		\hspace{10 mm}
		\text{and}
		\hspace{10 mm}
		j_{c_{2}}^{}=  
		\bordermatrix{~\cr
			1&0\cr
			2&0\cr
			3&0\cr
			4&j_{c_2}\cr
			5&j_{c_2}\cr
			6&j_{c_2}},
	\end{gather}  
	respectively. Each of these flux can be given by any of the elementary reactions of the proofreading network by inserting steady state concentrations of intermediate species within it,
	$$j_{c_1}=k_1^R(e_{ss}^R)-k_{-1}^R(eR_{ss}) \hspace{1 mm}
	\text{and}\hspace{1 mm}
	j_{c_{2}}=k_1^W(e_{ss}^W)-k_{-1}^W(eW_{ss}).$$ 
	Both $j_{c_1}$ and $j_{c_2}$ have mathematical expression same as eq.\eqref{f2} with two different superscript `R' and `W' respectively for rate constants. 
	\subsection{Dissipation of The Network}
	\begin{figure*}[htb!]
		\begin{center}
			\includegraphics[width=\textwidth]{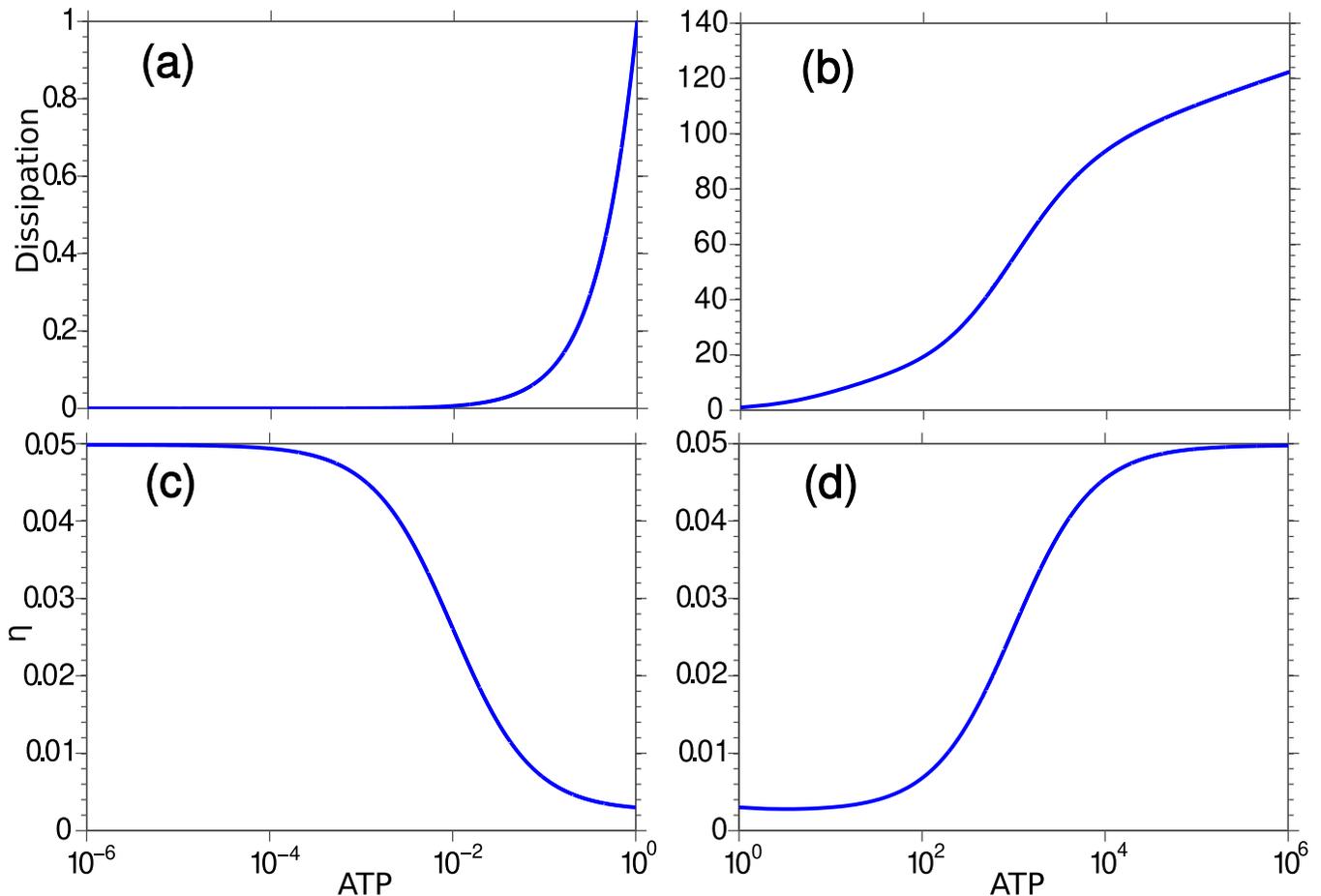}
			\caption{\label{fig:9} In \cref{fig:9}(a) and \cref{fig:9}(c), dissipation and error rate of the kinetic proofreading for the range of ATP concentration $10^{-6}-10^0$ is shown. As error rate deceases over the higher end of the concentration, dissipation increases. The dissipation and error rate of the kinetic proofreading for the higher range of ATP concentration $10^{0}-10^6$ is illustrated in \cref{fig:9}(b) and \cref{fig:9}(d), respectively. In this range of the concentration, dissipation qualitatively reflects the profile of the error rate. The rate constants of proofreading network are fixed at $k_1=5$,$k_2^0=1$ $k_3^R=1$, $k_{-1}^R=50$, $k_{-2}=10^{-3}$ and $k_{-3}=10^{-3}$.}
		\end{center}
	\end{figure*}
	We can represent total dissipation~\citep{Kondepudi2014ModernThermodynamics} of the reaction network using a flux-force relation. The overall steady state flux of kinetic proofreading network is acquired as the sum of fluxes of two emergent cycles, $J_{ss}=j_{c_1}+j_{c_2}$. Now the chemical force, reaction affinities\cite{Kondepudi2014ModernThermodynamics} acting along the cycles are represented by emergent affinities~\citep{Polettini2014IrreversibleLaws} which obeys following equation
	\begin{equation}
		\mu_{\epsilon}=c_{\epsilon}\ln{\frac{k_{\rho}}{k_{-\rho}}}z_{c}^{-S_{\rho}^c},
	\end{equation} 
	with $z_{c}$ and $S_{\rho}^c$ being concentrations and stoichiometric elements of chemostatted species, respectively. Therefore, the chemical affinities along $c_1$ and $c_2$ are $\mu_1=\ln{\frac{k_1^Rk_2^Rk_3^R}{k_{-1}^Rk_{-2}^Rk_{-3}^R}} $ and $\mu_2=\ln{\frac{k_1^Wk_2^Wk_3^W}{k_{-1}^Wk_{-2}^Wk_{-3}^W}}$. Any non-zero positive values of affinities $\mu_1$ and $\mu_2$ will drive the system out of equilibrium. Our assumptions for reaction rates here make sure that $\mu_1=\mu_2=\mu$. Thus, total dissipation of the proofreading network is obtained as, $ \dot{\Sigma}=J_{ss}\mu$. The second law of thermodynamics demands this dissipation to always be positive. 
	
	A change in ATP concentration from very low to a high level brings modulation in dissipation as illustrated in \cref{fig:9}. This change in the ATP concentration, keeping the concentration of ADP fixed at a certain level, implies a gradual increase in the ATP to ADP concentration ratio of the system. We have presented two breakdown figures \cref{fig:9}(a) and \cref{fig:9}(b) to understand the dissipation curve with respect to ATP concentration, and corresponding error variations are shown in \cref{fig:9}(c) and \cref{fig:9}(d). The range of ATP concentration variation on the left panel of \cref{fig:9} drives the system from its initial near-equilibrium state to nonequilibrium one. Due to the concentration change, the error in \cref{fig:9}(c) gradually decreases from near-equilibrium to the nonequilibrium regime. This error rate variation is expected from the error basin illustration in \cref{fig:1}(a) and corresponding discussion of \cref{error_basin}. Now from  \cref{fig:9}(c) and \cref{fig:9}(a), we can observe that dissipation becomes nonzero and increases monotonically as the error enters the nonequilibrium error regime from near-equilibrium. This dissipation curve reveals the trade-off nature of error and dissipation in this range of ATP concentration. For a much higher range of ATP concentration on the right panel of \cref{fig:9}, the error increases from its minimum value and approaches the equilibrium magnitude within the nonequilibrium environment in \cref{fig:9}(d). By comparing \cref{fig:9}(b) and \cref{fig:9}(d), it is observed that the corresponding dissipation curve qualitatively reflects the same dynamics as the error rate of the proofreading mechanism in this regime. A simultaneous increase in both error and dissipation in this range of ATP concentration hints at the lack of trade-off between error and dissipation under the nonequilibrium environment.
	\subsection{Semigrand Gibbs Free Energy}
	\begin{figure*}
		\begin{center}
			\includegraphics[width=\textwidth]{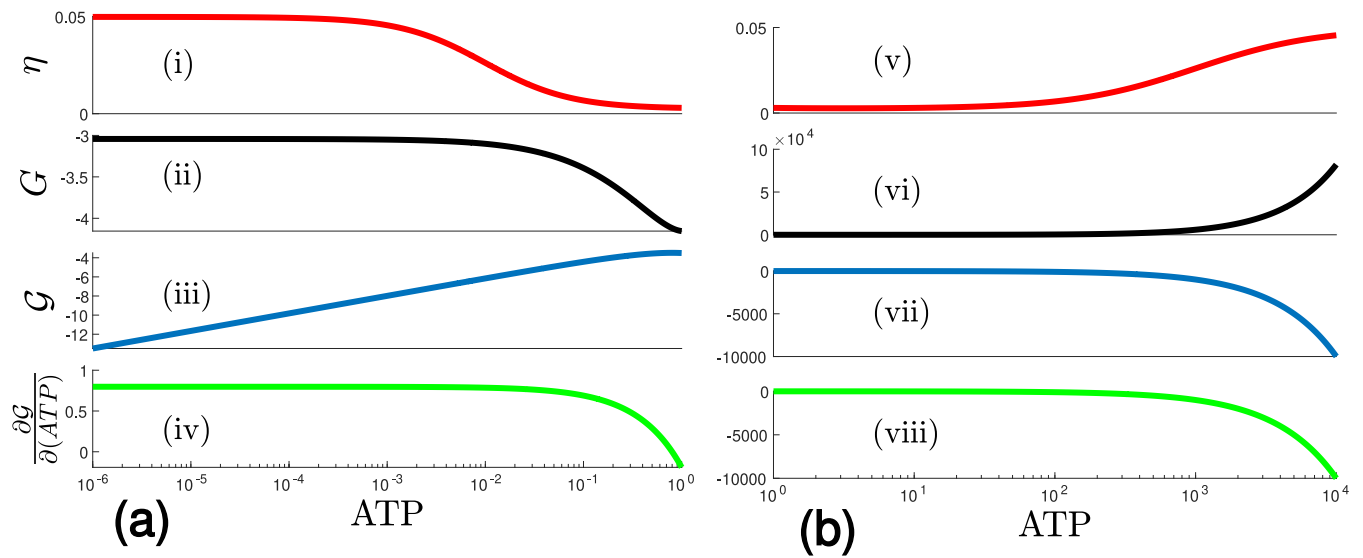}
			\caption{\label{fig:11} The variation of error rate, Gibbs free energy, semigrand Gibbs free energy and slope of semigrand Gibbs free energy for lower range of ATP concentration ($10^{-6}-10^0$)  is plotted in (i), (ii), (iii) and (iv) of \cref{fig:11}(a). Whereas response curve corresponding to higher range of ATP concentration is shown in (v), (vi), (vii) and (viii) of \cref{fig:11}(b). For both the cases, binding energy difference, $\Delta=3$ has been considered. The rate constants of proofreading network are fixed at $k_1=5$,$k_2^0=1$ $k_3^R=1$, $k_{-1}^R=50$, $k_{-2}=10^{-3}$ and $k_{-3}=10^{-3}$.}
		\end{center}
	\end{figure*}
	We now need to incorporate a proper nonequilibrium thermodynamic potential of the open system to analyze the energy expenditure of the proofreading network at the nonequilibrium regime. For this purpose, we have characterized each chemical species thermodynamically by chemical potential $\mu_{\sigma}$. The expression of the chemical potential is  $\mu_{\sigma}=\mu_{\sigma}^o+\ln{\frac{z_{\sigma}}{z_0}}$ with solvent concentration $z_0$ and standard-state chemical potential $\mu_{\sigma}^o$. The standard-state quantities with notation $`o'$ are taken at standard pressure $p=p^{o}$ and molecular concentration and chemical potential to set a baseline for substances. Moreover, a constant term like $\ln{z_{0}}$ can also be absorbed within $\mu_{\sigma}^o$ of the chemical potential. Thus, the nonequilibrium Gibbs free energy of a chemical reaction network can be written in terms of the chemical potential as~\citep{Fermi1956Thermodynamics} 
	\begin{equation}
		G=G_{0}+ \sum_{\sigma \neq 0}{(z_{\sigma}\mu_{\sigma}-z_{\sigma})}\label{neg}
	\end{equation}
	with $G_{0}=z_{0}\mu_{0}^o$. However, the proper entity to capture the energetics of the open system like the proofreading network is semigrand Gibbs free energy~\citep{Falasco2018InformationPatterns}. The semigrand Gibbs free energy of an open system is obtained from the nonequilibrium Gibbs free energy by a Legendre transformation~\citep{Rao2016NonequilibriumThermodynamics}
	\begin{equation}
		\mathcal{G}=G-\sum_{\lambda_b}{\mu_{\lambda_b}M_{\lambda_b}}. 
		\label{smgg}
	\end{equation}
	where $M_{\lambda_b}=\sum_{C_b}l_{C_b}^{{\lambda_b}^{-1}} L_{\lambda_b}$ represents moieties~\citep{Haraldsdttir2016IdentificationOC} exchanged between chemostats and system  through the external flow of the chemostatted species.
	
	The variation in the Gibbs free energy, the semigrand Gibbs free energy, and its slope due to change in ATP concentration over two different error rate regimes is presented in \cref{fig:11}(a) and \cref{fig:11}(b). The Gibbs free energy of the system exhibits similar qualitative characteristics as the error profile of the system for both the lower and higher range of ATP concentration in $(ii)$ and $(vi)$ of \cref{fig:11}(a) and \cref{fig:11}(b), respectively. For the concentration range corresponding to minimum error rate, when dissipation starts increasing from zero(\cref{fig:9}(a)), the Gibbs free energy lowers accordingly. However, for the higher range of ATP concentration, as the error grows after the plateau region of minimum error, the dissipation further enhances(\cref{fig:9}(b)) within a more strongly driven nonequilibrium environment, and the Gibbs free energy rapidly moves towards a very high positive value. The sum of nonequilibrium Gibbs free energy and dissipation of proofreading network quantifies the chemical work performed on the system~\citep{Rao2016NonequilibriumThermodynamics}. Thus we assert that the proofreading network performs with optimal error rate for small and finite chemical work. The chemical work diverges rapidly for the strongly driven nonequilibrium regime having lower accuracy.  
	
	Figure \ref{fig:11}(a) suggests that as more ATP accumulates in the system, the semigrand Gibbs free energy, $\mathcal{G}$, moves towards zero from a negative value. At the same time, the error rate goes towards a minimum. The corresponding semigrand Gibbs free energy slope remains constant initially and then goes from one to zero, as the error changes from the equilibrium value to the minimum one. For a comparatively higher ATP concentration in \cref{fig:11}(b), semigrand Gibbs free energy remains near zero over an extended range of ATP concentration and sustains the minimum error within a nonequilibrium environment. However, as the error curve deviates from its minimum value due to further change in ATP concentration, a sharp and vast change towards negative value can be seen in this semigrand Gibbs free energy profile. This massive change is also evident from the corresponding slope shown in $(viii)$ of \cref{fig:11}(b). The enormous magnitude of Gibbs free energy and semigrand Gibbs free energy in \cref{fig:11}(b) suggests that the system is far from equilibrium. Further, we conclude that the proofreading system operates at optimal accuracy and dissipation when semigrand Gibbs free energy has a value near zero.
	
	\section{\label{sec:VI} Error-Dissipation-Semigrand Gibbs Free Energy Profile and Performance of Proofreading Network} 
	\begin{figure*}
		\begin{center}
			\includegraphics[width=\textwidth]{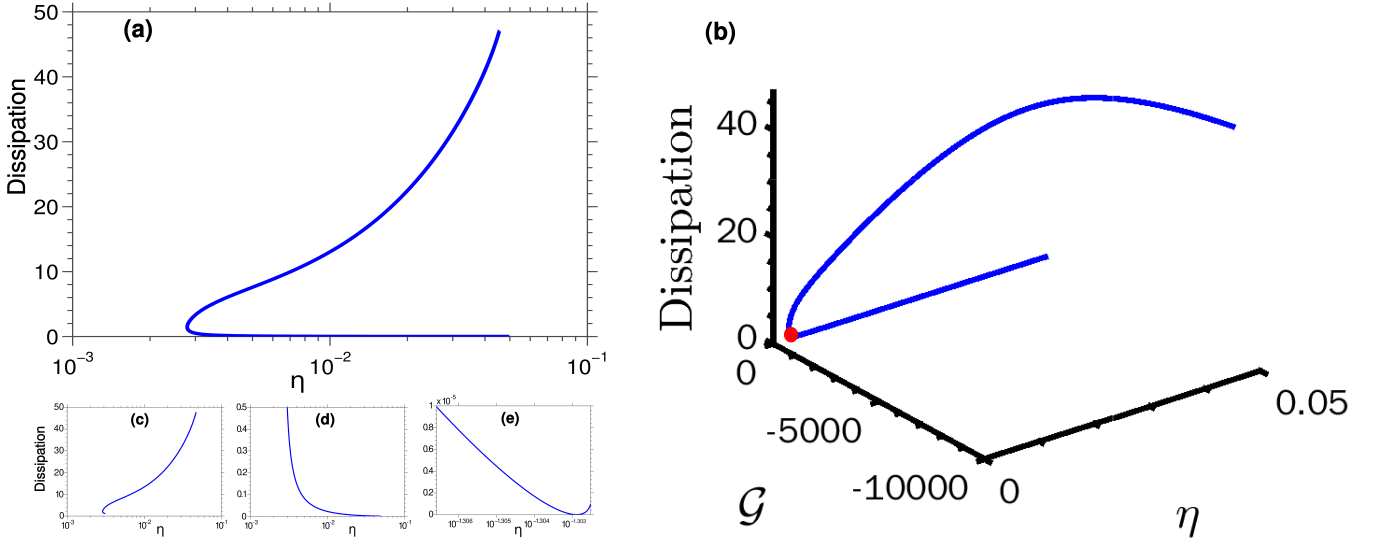}
			\caption{\label{fig:7} Dissipation with respect to the error rate is shown in \cref{fig:7}(a) for the range of ATP concentration, $10^{-6}-10^4$. The binding energy difference $\Delta=3$ has been considered. Figure \ref{fig:7}(e) corresponds to the regime($k_{-1} >> k_{+2}$) where the error rate is equal to the equilibrium error rate and dissipation is negligible. As the error rate decreases from equilibrium magnitude to the minimum one(in $k_{-1} \simeq k_{+2}$ regime), a sharp change in dissipation is shown in \cref{fig:7}(d). In \cref{fig:7}(c), both the error rate and dissipation increase (the $k_{-1} << k_{+2}$ regime). Figure \ref{fig:7}(b) illustrates the simultaneous variation of dissipation and semigrand Gibbs free energy concerning the error rate over the same ATP concentration range.}
		\end{center}
	\end{figure*}	 
	A trade-off between error and dissipation is present if any modification in a parameter cannot simultaneously improve both. However, this trade-off feature can be completely absent in other parts of the same process. We have investigated the trade-off nature of the dissipation and related error for different discrimination regimes that correspond to externally controlled ATP concentration variation. Association of Semigrand Gibbs free energy with this trade-off scenario is also demonstrated. Additionally, in the presence of distinct binding energy difference $\Delta$, we have unveiled the overall performance of the reaction network in terms of error and dissipation for the whole range of ATP concentration.   
	
	Here, \cref{fig:7}(a) illustrates a trade-off curve of error and dissipation with the error being exploited as an independent parameter. The figure suggests that dissipation is infinitesimally small till the minimum error is reached. From \cref{fig:7}(e), we can see the infinitesimal dissipation change near the equilibrium error regime. Then a sharp and finite increase in the dissipation near the minimum error rate can be seen from \cref{fig:7}(d). The fast-increasing part implies that as we further decrease the error rate, the dissipation inevitably increases as the virtue of dissipation-error trade-off in this regime. Then the error rate again approaches the magnitude of the equilibrium error(for a fixed lower $\Delta$) due to a further increase in the external driving force and dissipation monotonically increases owing to that external driving force in \cref{fig:7}(c). Thus, it is observed that the error value does not exclusively determine the dissipation as the dissipation can be different for the same error depending on the distance from the equilibrium. Further, \cref{fig:7}(b) exhibits the simultaneous variation of the dissipation and semigrand Gibbs free energy with the error rate and thus reveals the allocation of the proper energy content of the system to the error-dissipation trade-off scenarios. The red dot in the figure indicates the optimal working condition of the network having the minimal semigrand Gibbs free energy, dissipation, and error rate.   
	\begin{figure*}
		\begin{center}
			\includegraphics[width=\textwidth]{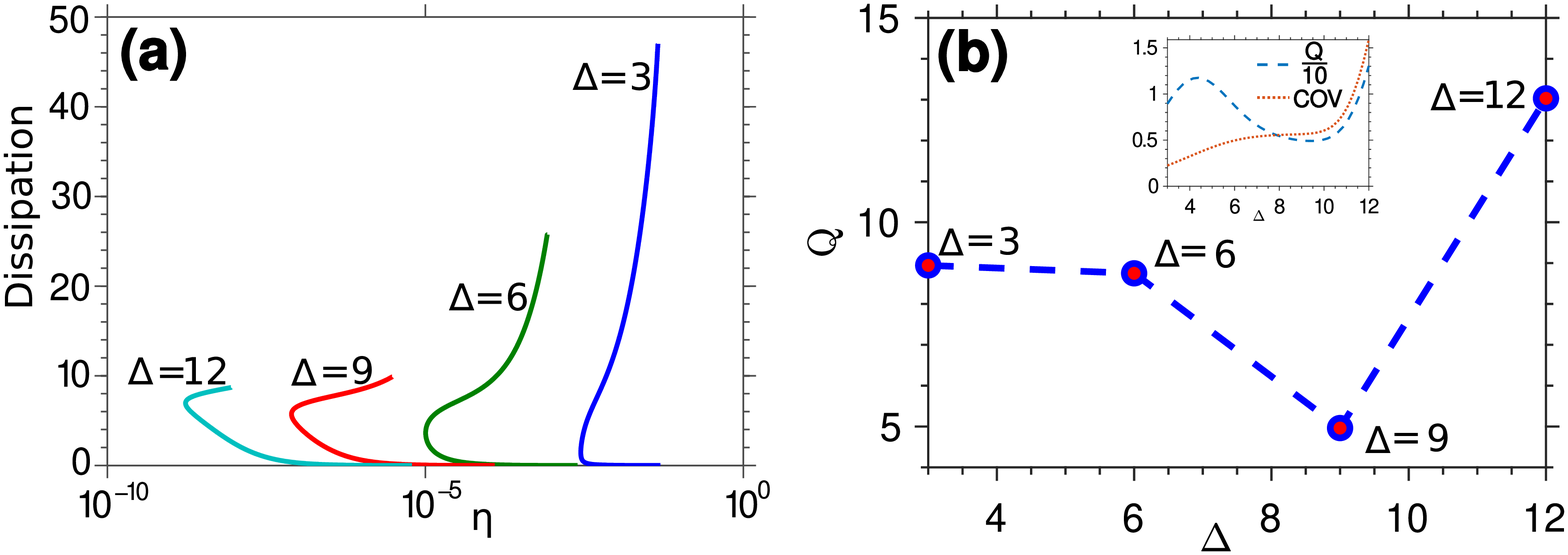}
			\caption{{\label{fig:8}} In \cref{fig:8}(a), the dissipation with respect to the error is shown for four different values of the binding energy difference, $\Delta(=3, 6, 9\hspace{1 mm} \text{and} \hspace{1 mm} 12)$. For all the cases, the range of ATP concentration varies from  $10^{-6}$ to $10^4$. All other parameters are fixed. Q value is defined as a product of the average dissipation and the coefficient of variation for the whole range of ATP concentration, as shown for $\Delta(=3, 6, 9\hspace{1 mm} \text{and} \hspace{1 mm} 12)$ in \cref{fig:8}(b). The lowest value of the Q  at $\Delta=9$ hints at comparatively good trade-off between error and dissipation for the whole range of the ATP concentration. Inset: Change in $Q$ factor and coefficients of variation(COV) due to the continuous variation of $\Delta$. The $Q$ factor profile is scaled by a factor $0.1$ to obtain a proper fit.} 
		\end{center} 
	\end{figure*}
	In \cref{fig:8}(a), the dissipation-error profile similar to \cref{fig:7}(a) is presented for four arbitrary magnitudes of the binding energy difference, $\Delta$. Although the error rate significantly decreases for higher $\Delta$ values, the maximum dissipation for the same range of ATP concentration decreases. The characteristic difference between the dissipation curves corresponding to higher and lower values of $\Delta$ is also evident from \cref{fig:8}(a). In the case of higher $\Delta$ values, the error rate of the proofreading scheme within the nonequilibrium environment gets stuck at a lower value for a much wider parameter space of ATP concentration. Thus, a more rapidly increasing nonzero portion of the dissipation plot for $\Delta=3$ than the same part of $\Delta=12$ is observed owing to the nonequilibrium regime with no improvement in error rate over the equilibrium error for relatively lower $\Delta$ values. It is also important to note from \cref{fig:8}(a) that the dissipation corresponding to the minimum error rate for higher $\Delta$ is greater than the one corresponding to the lower $\Delta$. As we have considered equal driving forces for different $\Delta$ values, these differences of dissipation curves are solely due to the different fluxes for distinct $\Delta$ values.  
	
	In a recent study~\citep{kpruncertainty}, the performance of the kinetic proofreading network representing biological operation has been studied with the aid of bound set by the thermodynamic uncertainty relation~\citep{BaratoSeifert}. However, we have analyzed the NESS of the kinetic proofreading network at a fixed time with a variation in ATP concentration and energetic discrimination. To quantify the performance of the network in terms of both error and dissipation in the presence of different binding energy differences, we have introduced a performance measuring entity at the NESS of the deterministic system for the variation of ATP concentration. For this purpose, we have set the error rate, $A_o={\eta}$ as the observable of our interest, and it changes with the concentration of ATP. Now at the NESS, we have denoted the mean and standard deviation associated with $A_o$ as $\left\langle A_o \right\rangle$ and $\sqrt{(\left\langle A_o^2 \right\rangle-\left\langle A_o \right\rangle^2)}$, respectively. Then, the coefficient of variation across the whole range ATP concentration in this case can be expressed as, $\epsilon=\frac{\sqrt{(\left\langle A_o^2 \right\rangle-\left\langle A_o \right\rangle^2)}}{\left\langle A_o \right\rangle}.$ Implementing the coefficient of variation is useful here as we have drastically different means corresponding to distinct $\Delta$ values. For a particular $\Delta$, we now define the performance measuring factor Q as a product of the average dissipation and the coefficient of variation, $Q=\left\langle \sigma \right\rangle \epsilon$. 
	
	In \cref{fig:8}(b), the factor Q decreases to a minimum value for $\Delta=9$ after having almost equal magnitude for $\Delta=3$ and $\Delta=6$. It is also evident from the \cref{fig:8}(b) that this $Q$ factor for $\Delta=12$ is even higher than the value corresponding to $\Delta=3$. The $Q$ factor and coefficients of variation corresponding to a continuous variation of $\Delta$ are illustrated in the inset of fig.~\cref{fig:8}(b). Despite the coefficients of variation corresponding to the error rate being lower for $\Delta=3$ and $\Delta=6$ than the other higher $\Delta$, a high average dissipation values for those two $\Delta$ push the corresponding $Q$ values above its value for  $\Delta=9$. However, for $\Delta=12$, even a much lower average dissipation is not enough to compensate for the higher coefficient of variation, and its' Q value is well above all other $\Delta$ values. Therefore, although average dissipation for the whole range of ATP concentration decreases monotonically with $\Delta$, the Q factor shows a non-monotonic behavior with $\Delta$. We assert that a lower Q value better adjusts error and dissipation to maintain optimal overall performance concerning different operating regimes over the whole ATP concentration range. Therefore, for a specific velocity and a fixed energy budget of the kinetic proofreading network, this Q term aids us in revealing a particular $\Delta$ that has optimal overall performance in terms of error and dissipation. 
	
	\section{Conclusions}
	We have described the kinetic proofreading through the substrate selection network from the standpoint of energetic discrimination and thereby exploited the chemical thermodynamic framework to analyze the operation of the network systematically. The main findings and strengths of our theoretical analysis include identification and characterization of the impact of the control parameters on the dynamic and thermodynamic entities of the network and the necessary formal development of proper energy content of a generic biological network. Further, our estimation of the nonequilibrium thermodynamic quantities aid in exploring the optimal regime of the proofreading in terms of error, dissipation, and semigrand Gibbs free energy. Finally, we have unveiled the overall performance of the proofreading network by defining a new entity, Q, in terms of the average dissipation and coefficient of variation for the whole range of a specific chemical fuel concentration for a particular binding energy difference. 
	
	Initially, it is shown that the proofreading kinetics can have different discrimination regimes depending on the relative strength between the pseudo-first-order rate constants. The association of the chemostatted species concentration with these pseudo-first-order rate constants makes it possible to access all different discrimination regimes at will by the concentration control of corresponding chemostatted species. Furthermore, the error basin found by changing the relative strength between rate constants experiences a modification depending on the magnitude of binding energy difference, $\Delta$.  Therefore, without changing the hardcore network structure, one can shift or modify the error regimes by the magnitude of the binding energy difference. This dependence on the binding energy magnitudes will also be vital if a family of competing substrates is present in a biological system. Further, in a biomolecular system, the impact of enhanced binding discrimination proves to be decisive in maintaining the system's operation close to the thermodynamic uncertainty relation~\citep{BaratoSeifert} bound or further from the bound~\citep{kpruncertainty}.

	By changing the concentration of chemical fuel, ATP, we have demonstrated that dissipation has the same qualitative variation as the error rate curve only for a higher concentration. On the contrary, the nonequilibrium Gibbs free energy of the system exhibits similar qualitative characteristics as the error profile of the system for both the lower and higher range of the concentration. The variation of these two quantities in different discrimination regimes establishes a relation between the chemical work and the error rate. Besides the conventional nonequilibrium Gibbs free energy, the semigrand Gibbs free energy provides important intuition about the network. It has been identified that the optimal network accuracy and dissipation are associated with near-zero semigrand Gibbs free energy. This finding in a simple proofreading network can be crucial and advantageous for searching the optimal specificity and dissipation regime in more complex and general proofreading networks and copying schemes~\citep{murugan2012, Murugan12034, wong2018}. This concentration control approach will be useful to estimate the energy cost of general biochemical feedback control mechanisms~\citep{Qianprl, Lan2012TheET}. 
	
	Additionally, the binding energy difference resulting in a lower Q value can be recommended for a certain purpose within a specific ATP concentration range. Thus, the Q factor can be an effective entity to be measured in synthetic biochemical networks to achieve a certain target with a concentration control. Indeed, the constraints and purposes of synthetic and biological engineered systems~\citep{salis2009automated, della2020engineering, pilsl2020optoribogenetic} are different from their natural counterparts. 
	
	Controlling discriminatory proofreading regimes and associated energetics through the concentrations of certain species and magnitude of binding energy difference would have crucial implications in designing more efficient synthetic biological architectures. This investigation can even be relevant for systematic performance analysis of self-assembled synthetic system~\citep{zhang2011dynamic, wei2012complex}. The regimes of trade-off and non-trade-off within the nonequilibrium environment hold the immense possibility to look into new control parameters and the concept of new mechanisms depending on the structure and energetics of the network, which can be captured via externally controlled chemostatted species. Knowledge of the proper energy consumption and storage of the network under continuous fuel consumption would also open the possibility and perspective for a more effective experimental design of tRNA Charging~\citep{Yamane2246}, translation by Ribosomes~\citep{Blanchard2004tRNASA}, and similar biological events. 
	\bibliography{ness}  

\begin{thebibliography}{43}%
\makeatletter
\providecommand \@ifxundefined [1]{%
 \@ifx{#1\undefined}
}%
\providecommand \@ifnum [1]{%
 \ifnum #1\expandafter \@firstoftwo
 \else \expandafter \@secondoftwo
 \fi
}%
\providecommand \@ifx [1]{%
 \ifx #1\expandafter \@firstoftwo
 \else \expandafter \@secondoftwo
 \fi
}%
\providecommand \natexlab [1]{#1}%
\providecommand \enquote  [1]{``#1''}%
\providecommand \bibnamefont  [1]{#1}%
\providecommand \bibfnamefont [1]{#1}%
\providecommand \citenamefont [1]{#1}%
\providecommand \href@noop [0]{\@secondoftwo}%
\providecommand \href [0]{\begingroup \@sanitize@url \@href}%
\providecommand \@href[1]{\@@startlink{#1}\@@href}%
\providecommand \@@href[1]{\endgroup#1\@@endlink}%
\providecommand \@sanitize@url [0]{\catcode `\\12\catcode `\$12\catcode
  `\&12\catcode `\#12\catcode `\^12\catcode `\_12\catcode `\%12\relax}%
\providecommand \@@startlink[1]{}%
\providecommand \@@endlink[0]{}%
\providecommand \url  [0]{\begingroup\@sanitize@url \@url }%
\providecommand \@url [1]{\endgroup\@href {#1}{\urlprefix }}%
\providecommand \urlprefix  [0]{URL }%
\providecommand \Eprint [0]{\href }%
\providecommand \doibase [0]{https://doi.org/}%
\providecommand \selectlanguage [0]{\@gobble}%
\providecommand \bibinfo  [0]{\@secondoftwo}%
\providecommand \bibfield  [0]{\@secondoftwo}%
\providecommand \translation [1]{[#1]}%
\providecommand \BibitemOpen [0]{}%
\providecommand \bibitemStop [0]{}%
\providecommand \bibitemNoStop [0]{.\EOS\space}%
\providecommand \EOS [0]{\spacefactor3000\relax}%
\providecommand \BibitemShut  [1]{\csname bibitem#1\endcsname}%
\let\auto@bib@innerbib\@empty
\bibitem [{\citenamefont {Hopfield}(1974)}]{Hopfield4135}%
  \BibitemOpen
  \bibfield  {author} {\bibinfo {author} {\bibfnamefont {J.~J.}\ \bibnamefont
  {Hopfield}},\ }\bibfield  {title} {\bibinfo {title} {Kinetic proofreading: A
  new mechanism for reducing errors in biosynthetic processes requiring high
  specificity},\ }\href {https://doi.org/10.1073/pnas.71.10.4135} {\bibfield
  {journal} {\bibinfo  {journal} {Proceedings of the National Academy of
  Sciences}\ }\textbf {\bibinfo {volume} {71}},\ \bibinfo {pages} {4135}
  (\bibinfo {year} {1974})},\ \Eprint
  {https://arxiv.org/abs/https://www.pnas.org/content/71/10/4135.full.pdf}
  {https://www.pnas.org/content/71/10/4135.full.pdf} \BibitemShut {NoStop}%
\bibitem [{\citenamefont {Ninio}(1975)}]{NINIO1975587}%
  \BibitemOpen
  \bibfield  {author} {\bibinfo {author} {\bibfnamefont {J.}~\bibnamefont
  {Ninio}},\ }\bibfield  {title} {\bibinfo {title} {Kinetic amplification of
  enzyme discrimination},\ }\href
  {https://doi.org/https://doi.org/10.1016/S0300-9084(75)80139-8} {\bibfield
  {journal} {\bibinfo  {journal} {Biochimie}\ }\textbf {\bibinfo {volume}
  {57}},\ \bibinfo {pages} {587 } (\bibinfo {year} {1975})}\BibitemShut
  {NoStop}%
\bibitem [{\citenamefont {Johnson}(1993)}]{johnson1993}%
  \BibitemOpen
  \bibfield  {author} {\bibinfo {author} {\bibfnamefont {K.~A.}\ \bibnamefont
  {Johnson}},\ }\bibfield  {title} {\bibinfo {title} {Conformational coupling
  in dna polymerase fidelity},\ }\href
  {https://doi.org/10.1146/annurev.bi.62.070193.003345} {\bibfield  {journal}
  {\bibinfo  {journal} {Annual Review of Biochemistry}\ }\textbf {\bibinfo
  {volume} {62}},\ \bibinfo {pages} {685} (\bibinfo {year} {1993})},\ \bibinfo
  {note} {pMID: 7688945},\ \Eprint
  {https://arxiv.org/abs/https://doi.org/10.1146/annurev.bi.62.070193.003345}
  {https://doi.org/10.1146/annurev.bi.62.070193.003345} \BibitemShut {NoStop}%
\bibitem [{\citenamefont {Schaaper}(1993)}]{SCHAAPER199323762}%
  \BibitemOpen
  \bibfield  {author} {\bibinfo {author} {\bibfnamefont {R.}~\bibnamefont
  {Schaaper}},\ }\bibfield  {title} {\bibinfo {title} {Base selection,
  proofreading, and mismatch repair during dna replication in escherichia
  coli.},\ }\href
  {https://doi.org/https://doi.org/10.1016/S0021-9258(20)80446-3} {\bibfield
  {journal} {\bibinfo  {journal} {Journal of Biological Chemistry}\ }\textbf
  {\bibinfo {volume} {268}},\ \bibinfo {pages} {23762} (\bibinfo {year}
  {1993})}\BibitemShut {NoStop}%
\bibitem [{\citenamefont {Blank}\ \emph {et~al.}(1986)\citenamefont {Blank},
  \citenamefont {Gallant}, \citenamefont {Burgess},\ and\ \citenamefont
  {Loeb}}]{Blank}%
  \BibitemOpen
  \bibfield  {author} {\bibinfo {author} {\bibfnamefont {A.}~\bibnamefont
  {Blank}}, \bibinfo {author} {\bibfnamefont {J.~A.}\ \bibnamefont {Gallant}},
  \bibinfo {author} {\bibfnamefont {R.~R.}\ \bibnamefont {Burgess}},\ and\
  \bibinfo {author} {\bibfnamefont {L.~A.}\ \bibnamefont {Loeb}},\ }\bibfield
  {title} {\bibinfo {title} {An rna polymerase mutant with reduced accuracy of
  chain elongation},\ }\href {https://doi.org/10.1021/bi00368a013} {\bibfield
  {journal} {\bibinfo  {journal} {Biochemistry}\ }\textbf {\bibinfo {volume}
  {25}},\ \bibinfo {pages} {5920} (\bibinfo {year} {1986})},\ \bibinfo {note}
  {pMID: 3098280},\ \Eprint
  {https://arxiv.org/abs/https://doi.org/10.1021/bi00368a013}
  {https://doi.org/10.1021/bi00368a013} \BibitemShut {NoStop}%
\bibitem [{\citenamefont {Gout}\ \emph {et~al.}(2017)\citenamefont {Gout},
  \citenamefont {Li}, \citenamefont {Fritsch}, \citenamefont {Li},
  \citenamefont {Haroon}, \citenamefont {Singh}, \citenamefont {Hua},
  \citenamefont {Fazelinia}, \citenamefont {Smith}, \citenamefont {Seeholzer},
  \citenamefont {Thomas}, \citenamefont {Lynch},\ and\ \citenamefont
  {Vermulst}}]{Goute}%
  \BibitemOpen
  \bibfield  {author} {\bibinfo {author} {\bibfnamefont {J.-F.}\ \bibnamefont
  {Gout}}, \bibinfo {author} {\bibfnamefont {W.}~\bibnamefont {Li}}, \bibinfo
  {author} {\bibfnamefont {C.}~\bibnamefont {Fritsch}}, \bibinfo {author}
  {\bibfnamefont {A.}~\bibnamefont {Li}}, \bibinfo {author} {\bibfnamefont
  {S.}~\bibnamefont {Haroon}}, \bibinfo {author} {\bibfnamefont
  {L.}~\bibnamefont {Singh}}, \bibinfo {author} {\bibfnamefont
  {D.}~\bibnamefont {Hua}}, \bibinfo {author} {\bibfnamefont {H.}~\bibnamefont
  {Fazelinia}}, \bibinfo {author} {\bibfnamefont {Z.}~\bibnamefont {Smith}},
  \bibinfo {author} {\bibfnamefont {S.}~\bibnamefont {Seeholzer}}, \bibinfo
  {author} {\bibfnamefont {K.}~\bibnamefont {Thomas}}, \bibinfo {author}
  {\bibfnamefont {M.}~\bibnamefont {Lynch}},\ and\ \bibinfo {author}
  {\bibfnamefont {M.}~\bibnamefont {Vermulst}},\ }\bibfield  {title} {\bibinfo
  {title} {The landscape of transcription errors in eukaryotic cells},\
  }\bibfield  {journal} {\bibinfo  {journal} {Science Advances}\ }\textbf
  {\bibinfo {volume} {3}},\ \href {https://doi.org/10.1126/sciadv.1701484}
  {10.1126/sciadv.1701484} (\bibinfo {year} {2017}),\ \Eprint
  {https://arxiv.org/abs/https://advances.sciencemag.org/content/3/10/e1701484.full.pdf}
  {https://advances.sciencemag.org/content/3/10/e1701484.full.pdf} \BibitemShut
  {NoStop}%
\bibitem [{\citenamefont {Zaher}\ and\ \citenamefont
  {Green}(2009)}]{ZAHER2009746}%
  \BibitemOpen
  \bibfield  {author} {\bibinfo {author} {\bibfnamefont {H.~S.}\ \bibnamefont
  {Zaher}}\ and\ \bibinfo {author} {\bibfnamefont {R.}~\bibnamefont {Green}},\
  }\bibfield  {title} {\bibinfo {title} {Fidelity at the molecular level:
  Lessons from protein synthesis},\ }\href
  {https://doi.org/https://doi.org/10.1016/j.cell.2009.01.036} {\bibfield
  {journal} {\bibinfo  {journal} {Cell}\ }\textbf {\bibinfo {volume} {136}},\
  \bibinfo {pages} {746 } (\bibinfo {year} {2009})}\BibitemShut {NoStop}%
\bibitem [{\citenamefont {Qian}(2006)}]{Qian1}%
  \BibitemOpen
  \bibfield  {author} {\bibinfo {author} {\bibfnamefont {H.}~\bibnamefont
  {Qian}},\ }\bibfield  {title} {\bibinfo {title} {Open-system nonequilibrium
  steady state: Statistical thermodynamics, fluctuations, and chemical
  oscillations},\ }\href {https://doi.org/10.1021/jp061858z} {\bibfield
  {journal} {\bibinfo  {journal} {The Journal of Physical Chemistry B}\
  }\textbf {\bibinfo {volume} {110}},\ \bibinfo {pages} {15063} (\bibinfo
  {year} {2006})},\ \Eprint
  {https://arxiv.org/abs/https://doi.org/10.1021/jp061858z}
  {https://doi.org/10.1021/jp061858z} \BibitemShut {NoStop}%
\bibitem [{\citenamefont {Qian}(2007)}]{Qian2}%
  \BibitemOpen
  \bibfield  {author} {\bibinfo {author} {\bibfnamefont {H.}~\bibnamefont
  {Qian}},\ }\bibfield  {title} {\bibinfo {title} {Phosphorylation energy
  hypothesis: Open chemical systems and their biological functions},\ }\href
  {https://doi.org/10.1146/annurev.physchem.58.032806.104550} {\bibfield
  {journal} {\bibinfo  {journal} {Annual Review of Physical Chemistry}\
  }\textbf {\bibinfo {volume} {58}},\ \bibinfo {pages} {113} (\bibinfo {year}
  {2007})},\ \Eprint
  {https://arxiv.org/abs/https://doi.org/10.1146/annurev.physchem.58.032806.104550}
  {https://doi.org/10.1146/annurev.physchem.58.032806.104550} \BibitemShut
  {NoStop}%
\bibitem [{\citenamefont {Ge}\ and\ \citenamefont {Qian}(2010)}]{qiandissi}%
  \BibitemOpen
  \bibfield  {author} {\bibinfo {author} {\bibfnamefont {H.}~\bibnamefont
  {Ge}}\ and\ \bibinfo {author} {\bibfnamefont {H.}~\bibnamefont {Qian}},\
  }\bibfield  {title} {\bibinfo {title} {Physical origins of entropy
  production, free energy dissipation, and their mathematical
  representations},\ }\href {https://doi.org/10.1103/PhysRevE.81.051133}
  {\bibfield  {journal} {\bibinfo  {journal} {Phys. Rev. E}\ }\textbf {\bibinfo
  {volume} {81}},\ \bibinfo {pages} {051133} (\bibinfo {year}
  {2010})}\BibitemShut {NoStop}%
\bibitem [{\citenamefont {Bennett}(1979)}]{BENNETT1979}%
  \BibitemOpen
  \bibfield  {author} {\bibinfo {author} {\bibfnamefont {C.~H.}\ \bibnamefont
  {Bennett}},\ }\bibfield  {title} {\bibinfo {title} {Dissipation-error
  tradeoff in proofreading},\ }\href
  {https://doi.org/https://doi.org/10.1016/0303-2647(79)90003-0} {\bibfield
  {journal} {\bibinfo  {journal} {Biosystems}\ }\textbf {\bibinfo {volume}
  {11}},\ \bibinfo {pages} {85 } (\bibinfo {year} {1979})}\BibitemShut
  {NoStop}%
\bibitem [{\citenamefont {Murugan}\ \emph {et~al.}(2012)\citenamefont
  {Murugan}, \citenamefont {Huse},\ and\ \citenamefont
  {Leibler}}]{Murugan12034}%
  \BibitemOpen
  \bibfield  {author} {\bibinfo {author} {\bibfnamefont {A.}~\bibnamefont
  {Murugan}}, \bibinfo {author} {\bibfnamefont {D.~A.}\ \bibnamefont {Huse}},\
  and\ \bibinfo {author} {\bibfnamefont {S.}~\bibnamefont {Leibler}},\
  }\bibfield  {title} {\bibinfo {title} {Speed, dissipation, and error in
  kinetic proofreading},\ }\href {https://doi.org/10.1073/pnas.1119911109}
  {\bibfield  {journal} {\bibinfo  {journal} {Proceedings of the National
  Academy of Sciences}\ }\textbf {\bibinfo {volume} {109}},\ \bibinfo {pages}
  {12034} (\bibinfo {year} {2012})},\ \Eprint
  {https://arxiv.org/abs/https://www.pnas.org/content/109/30/12034.full.pdf}
  {https://www.pnas.org/content/109/30/12034.full.pdf} \BibitemShut {NoStop}%
\bibitem [{\citenamefont {Mallory}\ \emph {et~al.}(2019)\citenamefont
  {Mallory}, \citenamefont {Kolomeisky},\ and\ \citenamefont
  {Igoshin}}]{Mallory}%
  \BibitemOpen
  \bibfield  {author} {\bibinfo {author} {\bibfnamefont {J.~D.}\ \bibnamefont
  {Mallory}}, \bibinfo {author} {\bibfnamefont {A.~B.}\ \bibnamefont
  {Kolomeisky}},\ and\ \bibinfo {author} {\bibfnamefont {O.~A.}\ \bibnamefont
  {Igoshin}},\ }\bibfield  {title} {\bibinfo {title} {Trade-offs between error,
  speed, noise, and energy dissipation in biological processes with
  proofreading},\ }\href {https://doi.org/10.1021/acs.jpcb.9b03757} {\bibfield
  {journal} {\bibinfo  {journal} {The Journal of Physical Chemistry B}\
  }\textbf {\bibinfo {volume} {123}},\ \bibinfo {pages} {4718} (\bibinfo {year}
  {2019})},\ \bibinfo {note} {pMID: 31074999},\ \Eprint
  {https://arxiv.org/abs/https://doi.org/10.1021/acs.jpcb.9b03757}
  {https://doi.org/10.1021/acs.jpcb.9b03757} \BibitemShut {NoStop}%
\bibitem [{\citenamefont {Murugan}\ \emph {et~al.}(2014)\citenamefont
  {Murugan}, \citenamefont {Huse},\ and\ \citenamefont
  {Leibler}}]{murugan2012}%
  \BibitemOpen
  \bibfield  {author} {\bibinfo {author} {\bibfnamefont {A.}~\bibnamefont
  {Murugan}}, \bibinfo {author} {\bibfnamefont {D.~A.}\ \bibnamefont {Huse}},\
  and\ \bibinfo {author} {\bibfnamefont {S.}~\bibnamefont {Leibler}},\
  }\bibfield  {title} {\bibinfo {title} {Discriminatory proofreading regimes in
  nonequilibrium systems},\ }\href {https://doi.org/10.1103/PhysRevX.4.021016}
  {\bibfield  {journal} {\bibinfo  {journal} {Phys. Rev. X}\ }\textbf {\bibinfo
  {volume} {4}},\ \bibinfo {pages} {021016} (\bibinfo {year}
  {2014})}\BibitemShut {NoStop}%
\bibitem [{\citenamefont {Wong}\ \emph {et~al.}(2018)\citenamefont {Wong},
  \citenamefont {Amir},\ and\ \citenamefont {Gunawardena}}]{wong2018}%
  \BibitemOpen
  \bibfield  {author} {\bibinfo {author} {\bibfnamefont {F.}~\bibnamefont
  {Wong}}, \bibinfo {author} {\bibfnamefont {A.}~\bibnamefont {Amir}},\ and\
  \bibinfo {author} {\bibfnamefont {J.}~\bibnamefont {Gunawardena}},\
  }\bibfield  {title} {\bibinfo {title} {Energy-speed-accuracy relation in
  complex networks for biological discrimination},\ }\href
  {https://doi.org/10.1103/PhysRevE.98.012420} {\bibfield  {journal} {\bibinfo
  {journal} {Phys. Rev. E}\ }\textbf {\bibinfo {volume} {98}},\ \bibinfo
  {pages} {012420} (\bibinfo {year} {2018})}\BibitemShut {NoStop}%
\bibitem [{\citenamefont {Banerjee}\ \emph {et~al.}(2020)\citenamefont
  {Banerjee}, \citenamefont {Das},\ and\ \citenamefont {Gangopadhyay}}]{kbgg}%
  \BibitemOpen
  \bibfield  {author} {\bibinfo {author} {\bibfnamefont {K.}~\bibnamefont
  {Banerjee}}, \bibinfo {author} {\bibfnamefont {B.}~\bibnamefont {Das}},\ and\
  \bibinfo {author} {\bibfnamefont {G.}~\bibnamefont {Gangopadhyay}},\
  }\bibfield  {title} {\bibinfo {title} {The guiding role of dissipation in
  kinetic proofreading networks: Implications for protein synthesis},\ }\href
  {https://doi.org/10.1063/1.5144726} {\bibfield  {journal} {\bibinfo
  {journal} {The Journal of Chemical Physics}\ }\textbf {\bibinfo {volume}
  {152}},\ \bibinfo {pages} {111102} (\bibinfo {year} {2020})},\ \Eprint
  {https://arxiv.org/abs/https://doi.org/10.1063/1.5144726}
  {https://doi.org/10.1063/1.5144726} \BibitemShut {NoStop}%
\bibitem [{\citenamefont {Savageau}\ and\ \citenamefont
  {Freter}(1979)}]{Savageau1979EnergyCO}%
  \BibitemOpen
  \bibfield  {author} {\bibinfo {author} {\bibfnamefont {M.~A.}\ \bibnamefont
  {Savageau}}\ and\ \bibinfo {author} {\bibfnamefont {R.~R.}\ \bibnamefont
  {Freter}},\ }\bibfield  {title} {\bibinfo {title} {Energy cost of
  proofreading to increase fidelity of transfer ribonucleic acid
  aminoacylation.},\ }\href@noop {} {\bibfield  {journal} {\bibinfo  {journal}
  {Biochemistry}\ }\textbf {\bibinfo {volume} {18 16}},\ \bibinfo {pages}
  {3486} (\bibinfo {year} {1979})}\BibitemShut {NoStop}%
\bibitem [{\citenamefont {Mehta}\ and\ \citenamefont
  {Schwab}(2012)}]{Mehta17978}%
  \BibitemOpen
  \bibfield  {author} {\bibinfo {author} {\bibfnamefont {P.}~\bibnamefont
  {Mehta}}\ and\ \bibinfo {author} {\bibfnamefont {D.~J.}\ \bibnamefont
  {Schwab}},\ }\bibfield  {title} {\bibinfo {title} {Energetic costs of
  cellular computation},\ }\href {https://doi.org/10.1073/pnas.1207814109}
  {\bibfield  {journal} {\bibinfo  {journal} {Proceedings of the National
  Academy of Sciences}\ }\textbf {\bibinfo {volume} {109}},\ \bibinfo {pages}
  {17978} (\bibinfo {year} {2012})},\ \Eprint
  {https://arxiv.org/abs/https://www.pnas.org/content/109/44/17978.full.pdf}
  {https://www.pnas.org/content/109/44/17978.full.pdf} \BibitemShut {NoStop}%
\bibitem [{\citenamefont {Rao}\ and\ \citenamefont
  {Esposito}(2016)}]{Rao2016NonequilibriumThermodynamics}%
  \BibitemOpen
  \bibfield  {author} {\bibinfo {author} {\bibfnamefont {R.}~\bibnamefont
  {Rao}}\ and\ \bibinfo {author} {\bibfnamefont {M.}~\bibnamefont {Esposito}},\
  }\bibfield  {title} {\bibinfo {title} {Nonequilibrium thermodynamics of
  chemical reaction networks: Wisdom from stochastic thermodynamics},\ }\href
  {https://doi.org/10.1103/PhysRevX.6.041064} {\bibfield  {journal} {\bibinfo
  {journal} {Phys. Rev. X}\ }\textbf {\bibinfo {volume} {6}},\ \bibinfo {pages}
  {041064} (\bibinfo {year} {2016})}\BibitemShut {NoStop}%
\bibitem [{\citenamefont {Falasco}\ \emph {et~al.}(2018)\citenamefont
  {Falasco}, \citenamefont {Rao},\ and\ \citenamefont
  {Esposito}}]{Falasco2018InformationPatterns}%
  \BibitemOpen
  \bibfield  {author} {\bibinfo {author} {\bibfnamefont {G.}~\bibnamefont
  {Falasco}}, \bibinfo {author} {\bibfnamefont {R.}~\bibnamefont {Rao}},\ and\
  \bibinfo {author} {\bibfnamefont {M.}~\bibnamefont {Esposito}},\ }\bibfield
  {title} {\bibinfo {title} {{Information Thermodynamics of Turing Patterns}},\
  }\href {https://doi.org/10.1103/PhysRevLett.121.108301} {\bibfield  {journal}
  {\bibinfo  {journal} {Physical Review Letters}\ }\textbf {\bibinfo {volume}
  {121}},\ \bibinfo {pages} {108301} (\bibinfo {year} {2018})}\BibitemShut
  {NoStop}%
\bibitem [{\citenamefont {Avanzini}\ and\ \citenamefont
  {Esposito}(2021)}]{Avanzini2021ThermodynamicsOC}%
  \BibitemOpen
  \bibfield  {author} {\bibinfo {author} {\bibfnamefont {F.}~\bibnamefont
  {Avanzini}}\ and\ \bibinfo {author} {\bibfnamefont {M.}~\bibnamefont
  {Esposito}},\ }\bibfield  {title} {\bibinfo {title} {Thermodynamics of
  concentration vs flux control in chemical reaction networks},\ }\bibfield
  {journal} {\bibinfo  {journal} {The Journal of Chemical Physics}\ }\href
  {https://doi.org/10.1063/5.0076134} {10.1063/5.0076134} (\bibinfo {year}
  {2021})\BibitemShut {NoStop}%
\bibitem [{\citenamefont {Hartich}\ \emph {et~al.}(2015)\citenamefont
  {Hartich}, \citenamefont {Barato},\ and\ \citenamefont
  {Seifert}}]{Hartich_2015}%
  \BibitemOpen
  \bibfield  {author} {\bibinfo {author} {\bibfnamefont {D.}~\bibnamefont
  {Hartich}}, \bibinfo {author} {\bibfnamefont {A.~C.}\ \bibnamefont
  {Barato}},\ and\ \bibinfo {author} {\bibfnamefont {U.}~\bibnamefont
  {Seifert}},\ }\bibfield  {title} {\bibinfo {title} {Nonequilibrium sensing
  and its analogy to kinetic proofreading},\ }\href
  {https://doi.org/10.1088/1367-2630/17/5/055026} {\bibfield  {journal}
  {\bibinfo  {journal} {New Journal of Physics}\ }\textbf {\bibinfo {volume}
  {17}},\ \bibinfo {pages} {055026} (\bibinfo {year} {2015})}\BibitemShut
  {NoStop}%
\bibitem [{\citenamefont {Lan}\ \emph {et~al.}(2012)\citenamefont {Lan},
  \citenamefont {Sartori}, \citenamefont {Neumann}, \citenamefont {Sourjik},\
  and\ \citenamefont {Tu}}]{Lan2012TheET}%
  \BibitemOpen
  \bibfield  {author} {\bibinfo {author} {\bibfnamefont {G.}~\bibnamefont
  {Lan}}, \bibinfo {author} {\bibfnamefont {P.}~\bibnamefont {Sartori}},
  \bibinfo {author} {\bibfnamefont {S.}~\bibnamefont {Neumann}}, \bibinfo
  {author} {\bibfnamefont {V.}~\bibnamefont {Sourjik}},\ and\ \bibinfo {author}
  {\bibfnamefont {Y.}~\bibnamefont {Tu}},\ }\bibfield  {title} {\bibinfo
  {title} {The energy-speed-accuracy tradeoff in sensory adaptation},\
  }\href@noop {} {\bibfield  {journal} {\bibinfo  {journal} {Nature physics}\
  }\textbf {\bibinfo {volume} {8}},\ \bibinfo {pages} {422 } (\bibinfo {year}
  {2012})}\BibitemShut {NoStop}%
\bibitem [{\citenamefont {Polettini}\ and\ \citenamefont
  {Esposito}(2014)}]{Polettini2014IrreversibleLaws}%
  \BibitemOpen
  \bibfield  {author} {\bibinfo {author} {\bibfnamefont {M.}~\bibnamefont
  {Polettini}}\ and\ \bibinfo {author} {\bibfnamefont {M.}~\bibnamefont
  {Esposito}},\ }\bibfield  {title} {\bibinfo {title} {{Irreversible
  thermodynamics of open chemical networks. I. Emergent cycles and broken
  conservation laws}},\ }\bibfield  {journal} {\bibinfo  {journal} {Journal of
  Chemical Physics}\ }\textbf {\bibinfo {volume} {141}},\ \href
  {https://doi.org/10.1063/1.4886396} {10.1063/1.4886396} (\bibinfo {year}
  {2014})\BibitemShut {NoStop}%
\bibitem [{\citenamefont {Polettini}\ \emph {et~al.}(2016)\citenamefont
  {Polettini}, \citenamefont {Bulnes-Cuetara},\ and\ \citenamefont
  {Esposito}}]{conlawPolettini}%
  \BibitemOpen
  \bibfield  {author} {\bibinfo {author} {\bibfnamefont {M.}~\bibnamefont
  {Polettini}}, \bibinfo {author} {\bibfnamefont {G.}~\bibnamefont
  {Bulnes-Cuetara}},\ and\ \bibinfo {author} {\bibfnamefont {M.}~\bibnamefont
  {Esposito}},\ }\bibfield  {title} {\bibinfo {title} {Conservation laws and
  symmetries in stochastic thermodynamics},\ }\href
  {https://doi.org/10.1103/PhysRevE.94.052117} {\bibfield  {journal} {\bibinfo
  {journal} {Phys. Rev. E}\ }\textbf {\bibinfo {volume} {94}},\ \bibinfo
  {pages} {052117} (\bibinfo {year} {2016})}\BibitemShut {NoStop}%
\bibitem [{\citenamefont {McKeithan}(1995)}]{McKeithan5042}%
  \BibitemOpen
  \bibfield  {author} {\bibinfo {author} {\bibfnamefont {T.~W.}\ \bibnamefont
  {McKeithan}},\ }\bibfield  {title} {\bibinfo {title} {Kinetic proofreading in
  t-cell receptor signal transduction},\ }\href
  {https://doi.org/10.1073/pnas.92.11.5042} {\bibfield  {journal} {\bibinfo
  {journal} {Proceedings of the National Academy of Sciences}\ }\textbf
  {\bibinfo {volume} {92}},\ \bibinfo {pages} {5042} (\bibinfo {year}
  {1995})},\ \Eprint
  {https://arxiv.org/abs/https://www.pnas.org/content/92/11/5042.full.pdf}
  {https://www.pnas.org/content/92/11/5042.full.pdf} \BibitemShut {NoStop}%
\bibitem [{\citenamefont {Banerjee}\ \emph {et~al.}(2017)\citenamefont
  {Banerjee}, \citenamefont {Kolomeisky},\ and\ \citenamefont
  {Igoshin}}]{Banerjee2017Accuracy}%
  \BibitemOpen
  \bibfield  {author} {\bibinfo {author} {\bibfnamefont {K.}~\bibnamefont
  {Banerjee}}, \bibinfo {author} {\bibfnamefont {A.~B.}\ \bibnamefont
  {Kolomeisky}},\ and\ \bibinfo {author} {\bibfnamefont {O.~A.}\ \bibnamefont
  {Igoshin}},\ }\bibfield  {title} {\bibinfo {title} {Accuracy of substrate
  selection by enzymes is controlled by kinetic discrimination.},\ }\href@noop
  {} {\bibfield  {journal} {\bibinfo  {journal} {The journal of physical
  chemistry letters}\ }\textbf {\bibinfo {volume} {8 7}},\ \bibinfo {pages}
  {1552} (\bibinfo {year} {2017})}\BibitemShut {NoStop}%
\bibitem [{\citenamefont {Kumar}\ and\ \citenamefont
  {Gangopadhyay}(2020)}]{pkgg}%
  \BibitemOpen
  \bibfield  {author} {\bibinfo {author} {\bibfnamefont {P.}~\bibnamefont
  {Kumar}}\ and\ \bibinfo {author} {\bibfnamefont {G.}~\bibnamefont
  {Gangopadhyay}},\ }\bibfield  {title} {\bibinfo {title} {Energetic and
  entropic cost due to overlapping of turing-hopf instabilities in the presence
  of cross diffusion},\ }\href {https://doi.org/10.1103/PhysRevE.101.042204}
  {\bibfield  {journal} {\bibinfo  {journal} {Phys. Rev. E}\ }\textbf {\bibinfo
  {volume} {101}},\ \bibinfo {pages} {042204} (\bibinfo {year}
  {2020})}\BibitemShut {NoStop}%
\bibitem [{\citenamefont {Kumar}\ and\ \citenamefont
  {Gangopadhyay}(2021)}]{pkgg2}%
  \BibitemOpen
  \bibfield  {author} {\bibinfo {author} {\bibfnamefont {P.}~\bibnamefont
  {Kumar}}\ and\ \bibinfo {author} {\bibfnamefont {G.}~\bibnamefont
  {Gangopadhyay}},\ }\bibfield  {title} {\bibinfo {title} {Nonequilibrium
  thermodynamics of glycolytic traveling wave: Benjamin-feir instability},\
  }\href {https://doi.org/10.1103/PhysRevE.104.014221} {\bibfield  {journal}
  {\bibinfo  {journal} {Phys. Rev. E}\ }\textbf {\bibinfo {volume} {104}},\
  \bibinfo {pages} {014221} (\bibinfo {year} {2021})}\BibitemShut {NoStop}%
\bibitem [{\citenamefont {Alberty}(2003)}]{Alberty2003ThermodynamicsReactions}%
  \BibitemOpen
  \bibfield  {author} {\bibinfo {author} {\bibfnamefont {R.~A.}\ \bibnamefont
  {Alberty}},\ }\href {https://doi.org/10.1002/0471332607} {\emph {\bibinfo
  {title} {{Thermodynamics of Biochemical Reactions}}}}\ (\bibinfo  {publisher}
  {John Wiley {\&} Sons, Inc.},\ \bibinfo {address} {Hoboken, NJ, USA},\
  \bibinfo {year} {2003})\BibitemShut {NoStop}%
\bibitem [{\citenamefont {Kondepudi}\ and\ \citenamefont
  {Prigogine}(2014)}]{Kondepudi2014ModernThermodynamics}%
  \BibitemOpen
  \bibfield  {author} {\bibinfo {author} {\bibfnamefont {D.}~\bibnamefont
  {Kondepudi}}\ and\ \bibinfo {author} {\bibfnamefont {I.}~\bibnamefont
  {Prigogine}},\ }\href {https://doi.org/10.1002/9781118698723} {\emph
  {\bibinfo {title} {{Modern Thermodynamics}}}}\ (\bibinfo  {publisher} {John
  Wiley {\&} Sons, Ltd},\ \bibinfo {address} {Chichester, UK},\ \bibinfo {year}
  {2014})\BibitemShut {NoStop}%
\bibitem [{\citenamefont {Fermi}(1956)}]{Fermi1956Thermodynamics}%
  \BibitemOpen
  \bibfield  {author} {\bibinfo {author} {\bibfnamefont {E.}~\bibnamefont
  {Fermi}},\ }\href
  {https://books.google.co.in/books/about/Thermodynamics.html?id=VEZ1ljsT3IwC&redir_esc=y}
  {\emph {\bibinfo {title} {{Thermodynamics}}}}\ (\bibinfo  {publisher} {Dover
  Publications},\ \bibinfo {year} {1956})\ p.\ \bibinfo {pages}
  {160}\BibitemShut {NoStop}%
\bibitem [{\citenamefont {Haraldsd{\'o}ttir}\ and\ \citenamefont
  {Fleming}(2016)}]{Haraldsdttir2016IdentificationOC}%
  \BibitemOpen
  \bibfield  {author} {\bibinfo {author} {\bibfnamefont {H.}~\bibnamefont
  {Haraldsd{\'o}ttir}}\ and\ \bibinfo {author} {\bibfnamefont {R.}~\bibnamefont
  {Fleming}},\ }\bibfield  {title} {\bibinfo {title} {Identification of
  conserved moieties in metabolic networks by graph theoretical analysis of
  atom transition networks},\ }\href@noop {} {\bibfield  {journal} {\bibinfo
  {journal} {PLoS Computational Biology}\ }\textbf {\bibinfo {volume} {12}}
  (\bibinfo {year} {2016})}\BibitemShut {NoStop}%
\bibitem [{\citenamefont {Pi\~neros}\ and\ \citenamefont
  {Tlusty}(2020)}]{kpruncertainty}%
  \BibitemOpen
  \bibfield  {author} {\bibinfo {author} {\bibfnamefont {W.~D.}\ \bibnamefont
  {Pi\~neros}}\ and\ \bibinfo {author} {\bibfnamefont {T.}~\bibnamefont
  {Tlusty}},\ }\bibfield  {title} {\bibinfo {title} {Kinetic proofreading and
  the limits of thermodynamic uncertainty},\ }\href
  {https://doi.org/10.1103/PhysRevE.101.022415} {\bibfield  {journal} {\bibinfo
   {journal} {Phys. Rev. E}\ }\textbf {\bibinfo {volume} {101}},\ \bibinfo
  {pages} {022415} (\bibinfo {year} {2020})}\BibitemShut {NoStop}%
\bibitem [{\citenamefont {Barato}\ and\ \citenamefont
  {Seifert}(2015)}]{BaratoSeifert}%
  \BibitemOpen
  \bibfield  {author} {\bibinfo {author} {\bibfnamefont {A.~C.}\ \bibnamefont
  {Barato}}\ and\ \bibinfo {author} {\bibfnamefont {U.}~\bibnamefont
  {Seifert}},\ }\bibfield  {title} {\bibinfo {title} {Thermodynamic uncertainty
  relation for biomolecular processes},\ }\href
  {https://doi.org/10.1103/PhysRevLett.114.158101} {\bibfield  {journal}
  {\bibinfo  {journal} {Phys. Rev. Lett.}\ }\textbf {\bibinfo {volume} {114}},\
  \bibinfo {pages} {158101} (\bibinfo {year} {2015})}\BibitemShut {NoStop}%
\bibitem [{\citenamefont {Qian}\ and\ \citenamefont {Reluga}(2005)}]{Qianprl}%
  \BibitemOpen
  \bibfield  {author} {\bibinfo {author} {\bibfnamefont {H.}~\bibnamefont
  {Qian}}\ and\ \bibinfo {author} {\bibfnamefont {T.~C.}\ \bibnamefont
  {Reluga}},\ }\bibfield  {title} {\bibinfo {title} {Nonequilibrium
  thermodynamics and nonlinear kinetics in a cellular signaling switch},\
  }\href {https://doi.org/10.1103/PhysRevLett.94.028101} {\bibfield  {journal}
  {\bibinfo  {journal} {Phys. Rev. Lett.}\ }\textbf {\bibinfo {volume} {94}},\
  \bibinfo {pages} {028101} (\bibinfo {year} {2005})}\BibitemShut {NoStop}%
\bibitem [{\citenamefont {Salis}\ \emph {et~al.}(2009)\citenamefont {Salis},
  \citenamefont {Mirsky},\ and\ \citenamefont {Voigt}}]{salis2009automated}%
  \BibitemOpen
  \bibfield  {author} {\bibinfo {author} {\bibfnamefont {H.~M.}\ \bibnamefont
  {Salis}}, \bibinfo {author} {\bibfnamefont {E.~A.}\ \bibnamefont {Mirsky}},\
  and\ \bibinfo {author} {\bibfnamefont {C.~A.}\ \bibnamefont {Voigt}},\
  }\bibfield  {title} {\bibinfo {title} {Automated design of synthetic ribosome
  binding sites to control protein expression},\ }\href@noop {} {\bibfield
  {journal} {\bibinfo  {journal} {Nature biotechnology}\ }\textbf {\bibinfo
  {volume} {27}},\ \bibinfo {pages} {946} (\bibinfo {year} {2009})}\BibitemShut
  {NoStop}%
\bibitem [{\citenamefont {Della~Corte}\ \emph {et~al.}(2020)\citenamefont
  {Della~Corte}, \citenamefont {van Beek}, \citenamefont {Syberg},
  \citenamefont {Schallmey}, \citenamefont {Tobola}, \citenamefont {Cormann},
  \citenamefont {Schlicker}, \citenamefont {Baumann}, \citenamefont {Krumbach},
  \citenamefont {Sokolowsky} \emph {et~al.}}]{della2020engineering}%
  \BibitemOpen
  \bibfield  {author} {\bibinfo {author} {\bibfnamefont {D.}~\bibnamefont
  {Della~Corte}}, \bibinfo {author} {\bibfnamefont {H.~L.}\ \bibnamefont {van
  Beek}}, \bibinfo {author} {\bibfnamefont {F.}~\bibnamefont {Syberg}},
  \bibinfo {author} {\bibfnamefont {M.}~\bibnamefont {Schallmey}}, \bibinfo
  {author} {\bibfnamefont {F.}~\bibnamefont {Tobola}}, \bibinfo {author}
  {\bibfnamefont {K.~U.}\ \bibnamefont {Cormann}}, \bibinfo {author}
  {\bibfnamefont {C.}~\bibnamefont {Schlicker}}, \bibinfo {author}
  {\bibfnamefont {P.~T.}\ \bibnamefont {Baumann}}, \bibinfo {author}
  {\bibfnamefont {K.}~\bibnamefont {Krumbach}}, \bibinfo {author}
  {\bibfnamefont {S.}~\bibnamefont {Sokolowsky}}, \emph {et~al.},\ }\bibfield
  {title} {\bibinfo {title} {Engineering and application of a biosensor with
  focused ligand specificity},\ }\href@noop {} {\bibfield  {journal} {\bibinfo
  {journal} {Nature Communications}\ }\textbf {\bibinfo {volume} {11}},\
  \bibinfo {pages} {1} (\bibinfo {year} {2020})}\BibitemShut {NoStop}%
\bibitem [{\citenamefont {Pilsl}\ \emph {et~al.}(2020)\citenamefont {Pilsl},
  \citenamefont {Morgan}, \citenamefont {Choukeife}, \citenamefont
  {M{\"o}glich},\ and\ \citenamefont {Mayer}}]{pilsl2020optoribogenetic}%
  \BibitemOpen
  \bibfield  {author} {\bibinfo {author} {\bibfnamefont {S.}~\bibnamefont
  {Pilsl}}, \bibinfo {author} {\bibfnamefont {C.}~\bibnamefont {Morgan}},
  \bibinfo {author} {\bibfnamefont {M.}~\bibnamefont {Choukeife}}, \bibinfo
  {author} {\bibfnamefont {A.}~\bibnamefont {M{\"o}glich}},\ and\ \bibinfo
  {author} {\bibfnamefont {G.}~\bibnamefont {Mayer}},\ }\bibfield  {title}
  {\bibinfo {title} {Optoribogenetic control of regulatory rna molecules},\
  }\href@noop {} {\bibfield  {journal} {\bibinfo  {journal} {Nature
  Communications}\ }\textbf {\bibinfo {volume} {11}},\ \bibinfo {pages} {1}
  (\bibinfo {year} {2020})}\BibitemShut {NoStop}%
\bibitem [{\citenamefont {Zhang}\ and\ \citenamefont
  {Seelig}(2011)}]{zhang2011dynamic}%
  \BibitemOpen
  \bibfield  {author} {\bibinfo {author} {\bibfnamefont {D.~Y.}\ \bibnamefont
  {Zhang}}\ and\ \bibinfo {author} {\bibfnamefont {G.}~\bibnamefont {Seelig}},\
  }\bibfield  {title} {\bibinfo {title} {Dynamic dna nanotechnology using
  strand-displacement reactions},\ }\href@noop {} {\bibfield  {journal}
  {\bibinfo  {journal} {Nature chemistry}\ }\textbf {\bibinfo {volume} {3}},\
  \bibinfo {pages} {103} (\bibinfo {year} {2011})}\BibitemShut {NoStop}%
\bibitem [{\citenamefont {Wei}\ \emph {et~al.}(2012)\citenamefont {Wei},
  \citenamefont {Dai},\ and\ \citenamefont {Yin}}]{wei2012complex}%
  \BibitemOpen
  \bibfield  {author} {\bibinfo {author} {\bibfnamefont {B.}~\bibnamefont
  {Wei}}, \bibinfo {author} {\bibfnamefont {M.}~\bibnamefont {Dai}},\ and\
  \bibinfo {author} {\bibfnamefont {P.}~\bibnamefont {Yin}},\ }\bibfield
  {title} {\bibinfo {title} {Complex shapes self-assembled from single-stranded
  dna tiles},\ }\href@noop {} {\bibfield  {journal} {\bibinfo  {journal}
  {Nature}\ }\textbf {\bibinfo {volume} {485}},\ \bibinfo {pages} {623}
  (\bibinfo {year} {2012})}\BibitemShut {NoStop}%
\bibitem [{\citenamefont {Yamane}\ and\ \citenamefont
  {Hopfield}(1977)}]{Yamane2246}%
  \BibitemOpen
  \bibfield  {author} {\bibinfo {author} {\bibfnamefont {T.}~\bibnamefont
  {Yamane}}\ and\ \bibinfo {author} {\bibfnamefont {J.~J.}\ \bibnamefont
  {Hopfield}},\ }\bibfield  {title} {\bibinfo {title} {Experimental evidence
  for kinetic proofreading in the aminoacylation of trna by synthetase},\
  }\href {https://doi.org/10.1073/pnas.74.6.2246} {\bibfield  {journal}
  {\bibinfo  {journal} {Proceedings of the National Academy of Sciences}\
  }\textbf {\bibinfo {volume} {74}},\ \bibinfo {pages} {2246} (\bibinfo {year}
  {1977})},\ \Eprint
  {https://arxiv.org/abs/https://www.pnas.org/content/74/6/2246.full.pdf}
  {https://www.pnas.org/content/74/6/2246.full.pdf} \BibitemShut {NoStop}%
\bibitem [{\citenamefont {Blanchard}\ \emph {et~al.}(2004)\citenamefont
  {Blanchard}, \citenamefont {Gonzalez}, \citenamefont {Kim}, \citenamefont
  {Chu},\ and\ \citenamefont {Puglisi}}]{Blanchard2004tRNASA}%
  \BibitemOpen
  \bibfield  {author} {\bibinfo {author} {\bibfnamefont {S.~C.}\ \bibnamefont
  {Blanchard}}, \bibinfo {author} {\bibfnamefont {R.~L.}\ \bibnamefont
  {Gonzalez}}, \bibinfo {author} {\bibfnamefont {H.~D.}\ \bibnamefont {Kim}},
  \bibinfo {author} {\bibfnamefont {S.}~\bibnamefont {Chu}},\ and\ \bibinfo
  {author} {\bibfnamefont {J.~D.}\ \bibnamefont {Puglisi}},\ }\bibfield
  {title} {\bibinfo {title} {trna selection and kinetic proofreading in
  translation},\ }\href@noop {} {\bibfield  {journal} {\bibinfo  {journal}
  {Nature Structural \&Molecular Biology}\ }\textbf {\bibinfo {volume} {11}},\
  \bibinfo {pages} {1008} (\bibinfo {year} {2004})}\BibitemShut {NoStop}%
\end{thebibliography}%
\end{document}